\documentclass[letterpaper,journal]{IEEEtran}
\usepackage{amsmath,amsfonts}
\usepackage{algorithmicx}
\usepackage{algcompatible}
\usepackage{algorithm}
\usepackage{subcaption}
\usepackage{array}

\usepackage{textcomp}
\usepackage{stfloats}
\usepackage{multirow}
\usepackage{fancyhdr}
\usepackage{url}
\usepackage{verbatim}
\usepackage[utf8]{inputenc}
\usepackage{booktabs}
\usepackage{tabularx}
\usepackage{graphicx}
\usepackage{hyperref}
\usepackage{xcolor}
\usepackage{makecell}
\usepackage[backend=biber,style=ieee,sorting=nty]{biblatex}
\makeatletter
\let\c@ALG@line\relax
\newcounter{ALG@line}
\makeatother
\begin{document}

%\title{DynamicAdaptiveClimb: Adaptive Cache Replacement with Dynamic Resizing$^*$\thanks{*A poster of an early version of this paper, by some of the authors of this paper, appeared in SYSTOR’19: Proceedings of the 12th ACM International Conference on Systems and Storage, Haifa, Israel (https://doi.org/10.1145/3319647.3325848).} \thanks{This work has also been awarded an official US Patent (https://patentcenter.uspto.gov/applications/16729695). }}

\title{DynamicAdaptiveClimb: Adaptive Cache Replacement with Dynamic Resizing$^{\ast}$\,$^{\|}$}
%\author{Daniel~Berend,
%        Shlomi~Dolev,
%        Sweta~Kumari,
%        Dhruv~Mishra,
%        Marina~Kogan-Sadetsky,
%        and~Archit~Somani% <- do not delete this line break comment
%\thanks{Daniel Berend, Shlomi Dolev, and Marina Kogan-Sadetsky are with the Department of Computer Science, Ben Gurion University, Beer-Sheva 8410501, Israel (e-mail: \{berend, dolev, sadetsky\}@cs.bgu.ac.il).}
%\thanks{Sweta Kumari, Dhruv Mishra, and Archit Somani are with the Department of Computer Science and Engineering, Shiv Nadar Institution of Eminence, Greater Noida 201314, India (e-mail: \{sweta.kumari, dm409, archit.somani\}@snu.edu.in).}
%}
\author{
  Daniel~Berend$^{1,2}$,
  Shlomi~Dolev$^1$,
  Sweta~Kumari$^3$,
  Dhruv~Mishra$^3$,
  Marina~Kogan-Sadetsky$^1$,
  and~Archit~Somani$^{3,\dagger}$
  \\
  Institute for the Theory of Computing, Ben Gurion University, Beer-Sheva, Israel$^1$
  \\
  Department of Mathematics, Ben Gurion University, Beer-Sheva, Israel$^2$
  \\
  \texttt{(berend, dolev, sadetsky)@bgu.ac.il}
  \\
  Department of Computer Science and Engineering, Shiv Nadar Institution of Eminence, India$^3$
  \\
  \texttt{(sweta.kumari, dm409, archit.somani)@snu.edu.in}
}

% The paper headers
%\markboth{Journal of \LaTeX\ Class Files,~Vol.~14, No.~8, August~2021}%
%{Shell \MakeLowercase{\textit{et al.}}: A Sample Article Using IEEEtran.cls for IEEE Journals}

%\IEEEpubid{0000--0000/00\$00.00~\copyright~2021 IEEE}
% Remember, if you use this, you must call \IEEEpubidadjcol in the second
% column for its text to clear the IEEEpubid mark.

\maketitle

% Manual footnotes using \renewcommand
\renewcommand{\thefootnote}{\fnsymbol{footnote}}

\footnotetext[1]{A poster of an early version of this paper, by some of the authors, appeared in SYSTOR’19: Proceedings of the 12th ACM International Conference on Systems and Storage, Haifa, Israel (\url{https://doi.org/10.1145/3319647.3325848}).}
\footnotetext[6]{This work has also been awarded an official US Patent (\url{https://patentcenter.uspto.gov/applications/16729695}).}
\footnotetext[2]{Corresponding author.}

% Reset footnote numbering to normal numbers if needed later
\renewcommand{\thefootnote}{\arabic{footnote}}

\begin{abstract}
Efficient cache management is critical for optimizing the system performance, and numerous caching mechanisms have been proposed, each exploring various insertion and eviction strategies. In this paper, we present \textit{AdaptiveClimb} and its extension, \textit{DynamicAdaptiveClimb}, two novel cache replacement policies that leverage lightweight, cache adaptation to outperform traditional approaches. Unlike classic Least Recently Used (LRU) and Incremental Rank Progress (CLIMB) policies, \textit{AdaptiveClimb} dynamically adjusts the promotion distance (\texttt{jump}) of the cached objects based on recent hit and miss patterns, requiring only a single tunable parameter and no per-item statistics. This enables rapid adaptation to changing access distributions while maintaining low overhead.
Building on this foundation, \textit{DynamicAdaptiveClimb} further enhances adaptability by automatically tuning the cache size in response to workload demands. Our comprehensive evaluation across a diverse set of real-world traces, including 1067 traces from 6 different datasets, demonstrates that \textit{DynamicAdaptiveClimb} consistently achieves substantial speedups and higher hit ratios compared to other state-of-the-art algorithms. In particular, our approach achieves up to a \textbf{29\% improvement in hit ratio} and a substantial reduction in miss penalties compared to the FIFO baseline. Furthermore, it outperforms the next-best contenders, \textit{AdaptiveClimb} and \textit{SIEVE}~\cite{yang2023sieve}, by approximately 10\% to 15\%, especially in environments characterized by fluctuating working set sizes. These results highlight the effectiveness of our approach in delivering efficient performance, making it well-suited for modern, dynamic caching environments.
\end{abstract}

%\thispagestyle{fancy}
%\fancyhf{} % clear header and footer
%\fancyfoot[C]{\href{https://patentcenter.uspto.gov/applications/16729695}{\textit{Official US Patent Link: https://patentcenter.uspto.gov/applications/16729695}}}

\section{Introduction}
In computing, a cache is a high-speed data storage layer that stores a subset of data, so that future requests for that data are served up faster. We would like the cache to store those data items that are most frequently accessed. In this work, we start from two well-known cache management algorithms, LRU\cite{lru} and CLIMB\cite{climb}, discuss their advantages and disadvantages, and suggest cache management algorithms that are able to combine the advantages of both. Let us first recall how the basic algorithms work. Both of them manage the order of the items in the cache. The idea is to place items, believed to be heavy hitters, at the top of the cache, and less frequent hitters closer to the bottom. 

\textbf{LRU:} Upon a request for some item $j$, which is not in the cache (cache miss), this item is inserted in the first position in the cache, all other items in the cache move down one position, and the item at the last position is evicted. If $j$ is in the cache (cache hit), say at position $i$, then it moves to the first position, while all items at positions $1$ to $i-1$ move down by one position.

\textbf{CLIMB:} Upon a request for item $j$, which is not in the cache, this item is placed at the last position in the cache, and the item at that position is evicted; when a cache hit occurs on an item at position $i$, this item exchanges places with the item at position $i-1$ (unless $i=1$, in which case there is no change).

LRU is known to be fast in adapting to changes in the distribution~\cite{lru}. CLIMB has been experimentally shown to have a higher hit ratio than LRU during periods of stable distribution~\cite{climb}. The reason is that CLIMB tends to keep the heavy hitters at top positions in the cache (delayed eviction), and hence, there is a smaller probability of them being evicted. This advantage comes at the expense of increased time to reach a steady state in comparison to LRU.

Let $N$ be the number of possible requests, $K$ the cache size, and $p_j$, $1 \leq j \leq N$, the probability of each item $j$ to be requested at each step. Under the Independent Requests (IR) model, the steady-state probabilities $\pi(\vec{\sigma})$ of each cache configuration $\vec{\sigma} = (\sigma_1, \sigma_2, \ldots, \sigma_K)$ are known for both algorithms~\cite{aven1987stochastic}:

\[
\pi^{LRU}(\vec{\sigma}) = \prod_{i=1}^K \frac{p_{\sigma_i}}{1 - \sum_{r=1}^{i-1} p_{\sigma_r}},
\]
\[
\pi^{CLIMB}(\vec{\sigma}) = C_1 \cdot \prod_{i=1}^K p_{\sigma_i}^{K-i+1},
\]
where $C_1$ is a normalization constant. We mention in passing that, although CLIMB has been shown experimentally to have a higher hit ratio than LRU under stable conditions, there is as yet no rigorous proof of this fact~\cite{aven1987stochastic}.

In the current paper, we introduce a unified framework and algorithms that combine the advantages of LRU and CLIMB. The idea is to make small changes in the cache when the situation is stable, and big changes when it is not. To this end, we introduce a \texttt{jump} size parameter. This parameter is the number of cells by which the current request is promoted in the cache, on its way from the bottom (or from the outside, in case of a cache miss) towards the top of the cache. From the point of view of the \texttt{jump} size, LRU and CLIMB may be viewed as two extremes. The \texttt{jump} size of LRU is $K$, which means a maximal change of the cache content due to each single request. On a cache hit, LRU promotes the current request by at most $K-1$ cells up, depending on its current position; in case of a miss, it \texttt{jumps} from outside the cache to the top, and in case of a hit, the item at location $i$ moves $i-1$ places to location $1$. This is the factor that makes LRU sensitive to data changes and adapt to changes quickly, compared with CLIMB. The \texttt{jump} size of CLIMB is $1$, which means a minimal change of the cache contents due to any single request. This \texttt{jump} size of CLIMB allows diminishing the influence of noise, namely the insertion of less frequent data items into the cache; a seldom-requested item inserted in the cache will most probably be removed from it before being requested again and promoted. CLIMB is better in gathering most frequently requested items in the cache during constant distribution periods. Our idea is to dynamically change the \texttt{jump} size so that it will fit both periods with frequent changes and periods with a constant distribution. Our algorithm achieves this goal by incrementing the \texttt{jump} size on cache misses and decrementing it on cache hits.

This longstanding dichotomy between adaptability and stability presents a fundamental challenge in cache management. Real-world access patterns are rarely static; they fluctuate, burst, and evolve. An ideal cache policy should not be forced to choose between responsiveness and robustness; rather, it should be capable of seamlessly transitioning between them based on observed workload characteristics.
To address this gap, we propose \textit{AdaptiveClimb}, a novel cache replacement algorithm that introduces a dynamic promotion mechanism based on a tunable parameter called \texttt{jump}. This parameter determines how far an item is moved toward the front of the cache upon access. Unlike static policies, \textit{AdaptiveClimb} adjusts the \texttt{jump} value in real time based on cache hits and misses, enabling the system to smoothly interpolate between LRU-like aggressiveness and CLIMB-like stability. This design allows \textit{AdaptiveClimb} to respond rapidly to changes when needed, while remaining resilient to noise during stable periods.
Building upon this foundation, we further introduce \textit{DynamicAdaptiveClimb}, which extends the self-tuning philosophy of \textit{AdaptiveClimb} by allowing the cache size itself to evolve in response to workload dynamics. By monitoring the behavior of both the full cache and its top segment using two promotion counters, the algorithm infers when the cache is either under-provisioned or over-provisioned and adjusts its size accordingly. This feature is especially valuable in multi-tenant systems or cloud environments, where cache is a shared or cost-sensitive resource.
Together, \textit{AdaptiveClimb} and \textit{DynamicAdaptiveClimb} represent a unified, lightweight framework for robust, real-time cache management. They require minimal parameter tuning, adapt organically to changing workloads, and offer strong empirical performance across diverse access patterns. By bridging the gap between LRU and CLIMB and incorporating dynamic cache sizing, our proposed algorithms advance the state-of-the-art in cache replacement and provide practical solutions for the evolving demands of high-performance systems.
%\newline
\smallskip

\noindent\textbf{The contributions of the paper are as follows:}
\begin{itemize}
    \item We propose \textit{AdaptiveClimb}, a self-tuning cache replacement algorithm that dynamically adjusts promotion distance based on access behavior.
    \item We introduce \textit{DynamicAdaptiveClimb}, which extends \textit{AdaptiveClimb} 
    by enabling automatic cache size adjustment with minimal overhead.
    \item We present a comprehensive analysis of both recent and state-of-the-art cache replacement algorithms to contextualize and benchmark our proposed approaches.
    \item Our approach forms a unified framework that bridges the adaptability of LRU and the stability of CLIMB.
    \item We demonstrate up to 29\% improvement in hit ratio over FIFO and 10–30\% gains over existing adaptive policies.
    \item Our algorithms are lightweight, requiring only two scalar variables, making them suitable for practical, resource-constrained environments.
\end{itemize}
\textbf{Roadmap.} Section~\ref{sec:relatedWork} reviews related work on cache replacement algorithms. Section~\ref{sec:sysModel} describes the system model used in this study. Section~\ref{sec:proposed} provides a detailed explanation of the proposed algorithms, AdaptiveClimb and DynamicAdaptiveClimb. Section~\ref{eval} presents a comprehensive performance evaluation of our algorithms in comparison with a diverse set of state-of-the-art cache replacement strategies. Finally, Section~\ref{sec:conclusion} concludes and outlines several directions for future work.

\section{Related Work}
\label{sec:relatedWork}
\begin{table*}[t]
\centering
\caption{Comparison of Popular Cache Replacement Algorithms}
\label{tab:cache-comparison}
\renewcommand{\arraystretch}{1.2}
\begin{tabularx}{\textwidth}{|p{2.6cm}|p{1cm}|p{3.3cm}|X|p{3.3cm}|}
\hline
\textbf{Algorithm} & \textbf{Year} & \textbf{Eviction Criteria} & \textbf{Use Case / Key Features} & \textbf{Simulator} \\
\hline
LRU~\cite{lru} & 1968 & Least Recently Used & Simple and widely deployed, fast to adapt & Synthetic Simulator \\
\hline
TwoQ~\cite{johnson1994twoq} & 1994 & Hot-Cold Queues & Separates new and frequent items in dual queues & Custom C++ Simulator \\
\hline
ARC~\cite{megiddo2003arc} & 2003 & Recency \& Frequency & Adapts between recency and frequency & Trace-Driven Simulator \\
\hline
LIRS~\cite{jiang2002lirs} & 2002 & Inter-Reference Recency & Replaces based on reuse distance; loop-friendly & Linux Kernel \\
\hline
CLOCK~\cite{jiang2005clockpro} & 2005 & Second-Chance Clock & Approximates LRU with less overhead & Trace Emulator \\
\hline
LHD~\cite{beckmann2018lhd} & 2016 & Reuse Distance Prediction & Predicts reuse distance for better decisions & RankCache \\
\hline
TinyLFU~\cite{eisman2017tinylfu} & 2017 & Frequency-based Admission & Probabilistic admission based on frequency sketch & Caffeine \\
\hline
B-LRU~\cite{twitterkv} & 2020 & Buffered Recency & Buffered LRU promotion reduces churn & Twitter Infra \\
\hline
SIEVE~\cite{yang2023sieve} & 2023 & Eviction Clustering & High throughput and simplicity & LibCacheSim \\
\hline
CACHEUS~\cite{bonomi2023cacheus} & 2023 & Online Learning & Learns eviction strategy from traffic patterns & Synthetic Simulator \\
\hline
GDSF-DBSCAN~\cite{bilal2023gdsfdbscan} & 2023 & Frequency \& Clustering & DBSCAN anomaly detection in cache eviction & Synthetic Simulator \\
\hline
Hyperbolic~\cite{torabi2023hrcache} & 2023 & Hazard Rate-Based & Targeted for edge devices with unstable patterns & C++ Trace-Driven Simulator \\
\hline
3L~\cite{pandey2024threeL} & 2024 & Multi-level Heuristic & Balances recency, frequency, and reusability & LibCacheSim \\
\hline
DFRC~\cite{sureshjani2024dfrc} & 2024 & Network-aware Policy & Drops easily retrievable content for NDN & Icarus \\
\hline
FLeeC~\cite{costa2024fleec} & 2024 & Lock-Free Concurrent & Optimized for concurrent workloads & MemCached \\
\hline
HashEvict~\cite{liu2024hashevict} & 2024 & Locality-sensitive Hashing & For key-value cache and transformer inference & Custom Simulator \\
\hline
ILRU~\cite{chen2025ilru} & 2025 & Incremental LRU & Robust for mobile/edge; low variance in hit rate & LibCacheSim \\
\hline
RAC~\cite{ahire2025rac} & 2025 & Random Adaptive & Reduces conflict misses via randomized placement & ChampSim \\
\hline
\textit{\textbf{AdaptiveClimb}} & \textbf{Proposed} & \textbf{Jump Size Tuning} & \textbf{Dynamically adapts promotion strategy for hits} & \textbf{LibCacheSim} \\
\hline
\textit{\textbf{DynamicAdaptiveClimb}} & \textbf{Proposed} & \textbf{Jump \& Cache Size Tuning} & \textbf{Cache size and promotion tuning for volatility} & \textbf{LibCacheSim} \\
\hline
\end{tabularx}
\end{table*}

In this section, we both analyze previous work in cache replacement and trace the evolution of algorithms over time. We categorize prior work into four primary paradigms: recency-based, frequency-based, recency-frequency hybrids, and modern adaptive or learning-based strategies. This structure enables a systematic exploration of the field. 
\begin{itemize}
\item Early recency-based approaches include LRU~\cite{lru}, CLOCK~\cite{corbato1968clock}, CLOCK-Pro~\cite{jiang2005clockpro}, and LIRS~\cite{jiang2002lirs}. 

\item Frequency-based methods such as LFU and its approximation TinyLFU~\cite{eisman2017tinylfu} represent a complementary class. 

\item Hybrid approaches balancing recency and frequency include ARC~\cite{megiddo2003arc}, CAR~\cite{bansal2004car}, MQ~\cite{zhou2004mq}, 2Q~\cite{johnson1994twoq}, LHD~\cite{beckmann2018lhd}, and B-LRU~\cite{bilal2023gdsfdbscan}. 

\item Modern adaptive or learning-based algorithms include SIEVE~\cite{yang2023sieve}, Cacheus~\cite{bonomi2023cacheus}, ILRU~\cite{chen2025ilru}, DFRC~\cite{sureshjani2024dfrc}, FLeeC~\cite{costa2024fleec}, 3L~\cite{pandey2024threeL}, RAC~\cite{ahire2025rac}, and HashEvict~\cite{liu2024hashevict}.
\end{itemize}

A number of prior works have conducted comprehensive evaluations and literature surveys on cache replacement policies. Yang et al.~\cite{twitterkv, fifo-all-you-need} benchmarked several eviction strategies at Twitter, revealing that simple policies like FIFO, when carefully tuned, can outperform sophisticated heuristics. Atikoglu et al.~\cite{atikoglu2012workload} and Huang et al.~\cite{huang2013facebook} studied workload behaviors in large-scale key-value stores and content delivery, respectively, emphasizing real-world skew, burstiness, and object variability. Yang et al.~\cite{twitterkv, yang2021segcache} analysed production caches and highlighted the limitations of classical approaches under scale. Additional comparative surveys by Liao et al.~\cite{lrb-original}, Zhang et al.~\cite{zhang2020survey}, and Bilal et al.~\cite{bilal2023survey} further underscore the need for hybrid, learning-based, and context-aware cache management solutions.

The design of cache replacement algorithms has remained a core research area for decades, evolving alongside the systems they serve. At the heart of this evolution lie two fundamental heuristics: recency and frequency. Early strategies like Least Recently Used (LRU)~\cite{lru} and CLOCK~\cite{corbato1968clock} approximate recency by evicting the least recently accessed items, offering simplicity and ease of hardware implementation. However, these policies often fail under bursty or non-stationary access patterns, where recentness does not imply future reuse.

To capture longer-term utility, frequency-based strategies such as LFU were introduced. Despite their theoretical appeal, LFU and its variants often struggle with cold-start bias and fail to adapt quickly to phase changes. TwoQ~\cite{johnson1994twoq} and MQ~\cite{zhou2004mq} attempt to reconcile these issues by maintaining separate buffers for short-term and long-term reuse, improving over LRU without incurring the full cost of LFU.

This foundational tension between recency and frequency sparked a line of research into hybrid and adaptive algorithms. Notably, LIRS~\cite{jiang2002lirs} redefined recency through inter-reference intervals and outperformed both LRU and LFU on typical workloads. ARC~\cite{megiddo2003arc} introduced a breakthrough in self-tuning policies, dynamically balancing recency and frequency through internal feedback mechanisms. CAR~\cite{bansal2004car} and CLOCK-Pro~\cite{jiang2005clockpro} extended these principles while preserving low computational overhead, making them viable for large-scale systems.

More recently, the shift toward cloud-scale, latency-sensitive, and multi-tenant systems has driven the need for efficient, low-overhead adaptability. In this context, TinyLFU~\cite{eisman2017tinylfu} offered a highly efficient admission control layer, filtering transient items using approximate frequency counters. LHD~\cite{beckmann2018lhd} proposed a probabilistic model of reuse distance to inform eviction, balancing prediction accuracy with computational cost. These algorithms mark a trend toward statistical learning as a replacement for hand-crafted eviction rules.

At the same time, researchers have revisited simplicity as a design principle. HillClimb~\cite{patel2021hillclimbalt} and SIEVE~\cite{yang2023sieve} are based on the argument that carefully designed heuristics, grounded in empirical reuse patterns, can outperform more complex methods. SIEVE, in particular, demonstrates that a deterministic eviction strategy with minimal metadata can scale efficiently and match or surpass policies that rely on learning or clustering.

In parallel, the rise of learning-based policies reflects growing interest in automatic adaptability. CACHEUS~\cite{bonomi2023cacheus} learns cache behavior online, adjusting to workload shifts in real time. GDSF-DBSCAN~\cite{bilal2023gdsfdbscan} introduces clustering to detect anomalies and manage outliers. HR-Cache~\cite{torabi2023hrcache} leverages hazard-rate models to cope with uncertainty in edge computing.

The emergence of domain-specific and hardware-constrained environments has further diversified the design space. For edge networks, DFRC~\cite{sureshjani2024dfrc} optimizes utility by evicting content that is rapidly retrievable from upstream caches. HashEvict~\cite{liu2024hashevict} targets transformer-based inference caches, using locality-sensitive hashing to manage key-value pairs. FLeeC~\cite{costa2024fleec} introduces lock-free concurrency for multi-threaded services, while 3L~\cite{pandey2024threeL} combines recency, frequency, and reusability across hierarchical cache layers.

In hardware-oriented contexts, ILRU~\cite{chen2025ilru} proposes an incremental caching model where items are partially cached based on access frequency, offering robustness for mobile and edge deployments. RAC~\cite{ahire2025rac} leverages randomized placement and associativity to reduce cache conflicts, offering an alternative to conventional index-based eviction.

Together, these developments reveal an important insight: no single heuristic suffices under the variety and volatility of modern workloads. Even advanced policies must now contend with trade-offs in responsiveness, overhead, transparency, and deployability.

Despite the proliferation of sophisticated approaches, many real-world deployments still rely on legacy strategies like LRU due to their simplicity, predictability, and ease of implementation. This gap between academic innovation and production deployment underscores a key challenge: how to design policies that are both adaptive and practical, capable of autonomously tuning behavior without sacrificing efficiency or transparency.

Our work builds on this observation by proposing two lightweight strategies, \textit{AdaptiveClimb} and \textit{DynamicAdaptiveClimb}, that reframe adaptability as a control-theoretic problem rather than a classification or prediction task. Instead of tracking frequency or computing scores, \textit{AdaptiveClimb} adjusts its promotion threshold via a single variable, \texttt{jump}, tuned by hit and miss feedback. This simple dynamic effectively interpolates between aggressive recency and conservative retention, sidestepping the complexity of LFU-style counters or learned models.

\textit{DynamicAdaptiveClimb} extends this logic by introducing cache resizing as a form of adaptability. Inspired by systems like Segcache~\cite{yang2021segcache} and Twitter's observations on workload variance~\cite{twitterkv}, it dynamically reallocates capacity based on spatial hit distributions, a feature largely absent in prior eviction strategies.

Positioned against existing research, our models offer a middle ground between the statistical sophistication of policies like LHD and CACHEUS and the pragmatic simplicity of LRU and SIEVE. Their design promotes low runtime cost, hardware friendliness, and stability under varying workloads, features often overlooked in policies optimized for theoretical performance but constrained in practical application.

A summary of the comparative analysis of these algorithms is done in Table \ref{tab:cache-comparison}.

\section{System Model Settings}
\label{sec:sysModel}
This section gives a brief foundation for the system model settings used for the algorithm proposal.

Suppose, in general, we have a list $R = \{r_1, r_2, \dots, r_N\}$ of $N$ possible different requests. Let $p_1 \geq p_2 \geq \dots \geq p_N$ be their (unknown) corresponding probabilities. The system contains a cache of size $K$ and a slower memory of size $ N-K$, where we assume $K < N$.

For each request $r$ and cache configuration $C$, the system must determine which configuration $C'$ and \texttt{jump} size to move to, depending on whether the request is a hit or a miss, and the current value of the \texttt{jump} size. 

On a cache \textit{miss}, the algorithm decides whether to insert the new item into the cache and, if so, which item to evict and how to update the cache contents. On a cache \textit{hit}, the algorithm may modify the position of the requested item $r$ (and necessarily adjust positions of other items accordingly) to reflect its updated access status.

\section{Proposed Algorithms}
\label{sec:proposed}
In this section, we introduce two novel cache replacement algorithms, \textit{AdaptiveClimb} and \textit{DynamicAdaptiveClimb}, that address the limitations of existing approaches by incorporating self-tuning promotion strategies and dynamic cache resizing to improve hit ratios under diverse workload conditions.
\subsection{AdaptiveClimb Algorithm}

The \textit{AdaptiveClimb} algorithm is a lightweight, adaptive cache replacement policy that introduces a single tunable parameter, \texttt{jump}, to govern item promotion within the cache. This scalar variable, which dynamically adjusts its value between 1 and $K$ (where $K$ is the cache size), controls how aggressively items are repositioned towards the front of the cache upon access. The algorithm is formally described in Algorithm~\ref{alg:adaptiveclimb}.
Unlike static algorithms such as LRU or LFU that rigidly prioritize either recency or frequency, \textit{AdaptiveClimb} responds fluidly to the recent access history by adjusting the magnitude of item promotion. Upon a cache hit \textit{(line no 1.2-1.5)}, the algorithm attempts to reduce the value of \texttt{jump}, thereby increasing the likelihood that frequently accessed items are promoted closer to the front of the cache. Conversely, on a cache miss \textit{(line no 1.10-1.13)}, \texttt{jump} is incremented, delaying promotion and increasing eviction pressure on infrequently accessed or newly introduced items.
The core idea behind this mechanism is to balance adaptability with stability: when workload patterns exhibit temporal locality, repeated accesses reduce the \texttt{jump} value and rapidly concentrate high-utility objects at the front. In contrast, during periods of high cache churn or low hit rates, the increasing \texttt{jump} value mitigates premature promotions of transient items, reducing cache pollution.
This simple yet effective feedback mechanism makes \textit{AdaptiveClimb} particularly suitable for workloads characterized by shifting access patterns. The complete operational flow is detailed in Algorithm~\ref{alg:adaptiveclimb}, which outlines how \texttt{jump} is updated \textit{(line no 1.4,1.12)} and how items are repositioned based on cache hits or misses.

\makeatletter
\setcounter{ALG@line}{0}
\renewcommand{\alglinenumber}[1]{1.\arabic{ALG@line}}
\makeatother
\begin{algorithm}[ht]
\caption{AdaptiveClimb($K$)}
\label{alg:adaptiveclimb}
\begin{algorithmic}[1]
\STATE \textbf{INITIALIZATION:} \texttt{jump} $\gets K$
\vspace{1mm}
\STATE \textbf{ON CACHE HIT} \texttt{cache}[$i$]
    \IF{\texttt{jump} $> 1$}
        \STATE \texttt{jump} $\gets$ \texttt{jump} $- 1$
    \ENDIF
    \IF{$i > 1$}
        \STATE Shift down elements between \texttt{cache}[$i - \texttt{jump}$] and \texttt{cache}[$i - 1$]
        \STATE \texttt{cache}[$i - \texttt{jump}$] $\gets$ \texttt{cache}[$i$]
    \ENDIF
\vspace{1mm}
\STATE \textbf{ON CACHE MISS} request $j$
    \IF{\texttt{jump} $< K$}
        \STATE \texttt{jump} $\gets$ \texttt{jump} $+ 1$
    \ENDIF
    \STATE Evict \texttt{cache}[$K$]
    \STATE Shift down elements between \texttt{cache}[$K - 1$] and \texttt{cache}[$K - \texttt{jump} + 1$]
    \STATE \texttt{cache}[$K - \texttt{jump} + 1$] $\gets j$
\end{algorithmic}
\end{algorithm}

\begin{figure}[!t]
    \centering
    % Subfigure 1 - Initial State
    \begin{subfigure}[t]{0.20\textwidth}
        \centering
        \includegraphics[width=\textwidth]{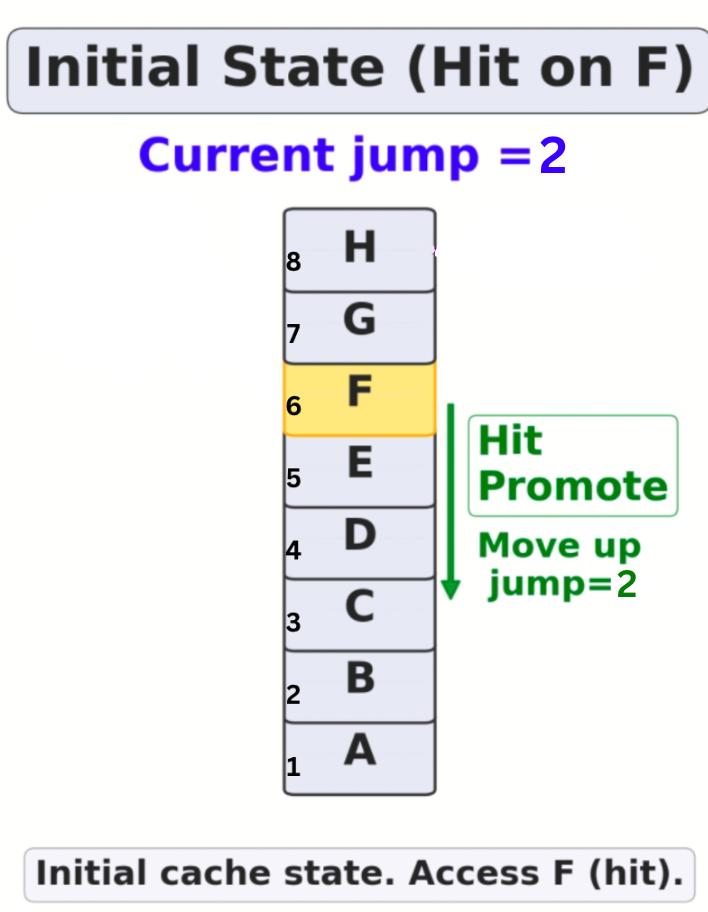}
        \caption{\textit{Initial State}}
        \label{fig:initial_hit}
    \end{subfigure}
    \hfill
    % Subfigure 2 - Final State
    \begin{subfigure}[t]{0.20\textwidth}
        \centering
        \includegraphics[width=\textwidth]{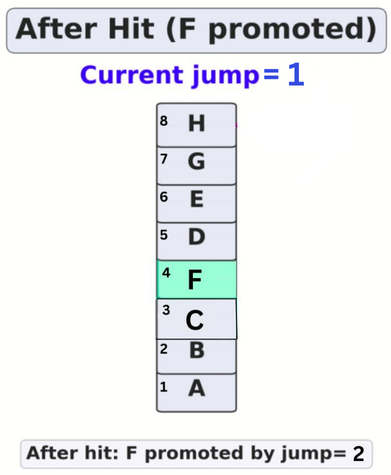}
        \caption{\textit{Final State}}
        \label{fig:final_hit}
    \end{subfigure}
    \caption{Cache states on hit under AdaptiveClimb.}
    \label{fig:adaptiveclimb_hit}
\end{figure}
Fig.~\ref{fig:adaptiveclimb_hit} illustrates the cache state transitions in the \textit{AdaptiveClimb} policy during a cache hit event. In Fig. \ref{fig:initial_hit}, the cache contains items labelled A through H, with the current \texttt{jump} parameter set to 2. When an access occurs on item F (a cache hit), the policy promotes F upward by two positions in the cache, as indicated by the green arrow and the ``Hit Promote'' annotation. Fig. \ref{fig:final_hit} shows the cache after the promotion, where F has moved up by two slots, and the rest of the items have shifted accordingly. 

\makeatletter
\setcounter{ALG@line}{0}
\renewcommand{\alglinenumber}[1]{2.\arabic{ALG@line}}
\makeatother
\begin{algorithm}[!t]
\caption{DynamicAdaptiveClimb($K, \varepsilon$)}
\label{alg:dynamicadaptiveclimb}
\begin{algorithmic}[1]
\STATE \texttt{jump} $\gets K$
\STATE \texttt{jump$'$} $\gets 0$
\vspace{1mm}
\STATE \textbf{ON CACHE HIT} \texttt{cache}[$i$]
    \IF{\texttt{jump} $> -K/2$}
        \STATE \texttt{jump} $\gets$ \texttt{jump} $- 1$
    \ENDIF
    \IF{$i \leq K/2$}
        \IF{\texttt{jump$'$} $> -K/2$}
            \STATE \texttt{jump$'$} $\gets$ \texttt{jump$'$} $- 1$
        \ENDIF
    \ELSE
        \IF{\texttt{jump$'$} $< 0$}
            \STATE \texttt{jump$'$} $\gets$ \texttt{jump$'$} $+ 1$
        \ENDIF
    \ENDIF
    \STATE \texttt{actualJump} $\gets \max\{1, \min\{\texttt{jump}, i - 1\}\}$
    \IF{$i > 1$}
        \STATE Shift down elements between \texttt{cache}[$i - \texttt{actualJump}$] and \texttt{cache}[$i - 1$]
        \STATE \texttt{cache}[$i - \texttt{actualJump}$] $\gets$ \texttt{cache}[$i$]
    \ENDIF
\vspace{1mm}
\STATE \textbf{ON CACHE MISS} request $j$
    \STATE \texttt{jump} $\gets$ \texttt{jump} $+ 1$
    \IF{\texttt{jump$'$} $< 0$}
        \STATE \texttt{jump$'$} $\gets$ \texttt{jump$'$} $+ 1$
    \ENDIF
    \STATE Evict \texttt{cache}[$K$]
    \STATE \texttt{actualJump} $\gets \max\{1, \min\{K - 1, \texttt{jump}\}\}$
    \STATE Shift down elements between \texttt{cache}[$K$] and \texttt{cache}[$K - \texttt{actualJump} + 1$]
    \STATE \texttt{cache}[$K - \texttt{actualJump} + 1$] $\gets j$
\vspace{1mm}
\IF{\texttt{jump} $== 0$}
    \STATE \texttt{jump$'$} $\gets 0$
\ENDIF
\IF{\texttt{jump} $== 2 \cdot K$}
    \STATE $K \gets 2 \cdot K$
\ENDIF
\IF{\texttt{jump} $== -K/2$ \textbf{AND} \texttt{jump$'$} $== -K/2 \cdot \varepsilon$}
    \STATE $K \gets K / 2$
\ENDIF
\end{algorithmic}
\end{algorithm}

\begin{figure}[!t]
    \centering
    % Subfigure 1 - Initial State
    \begin{subfigure}[t]{0.2\textwidth}
        \centering
        \includegraphics[width=\textwidth]{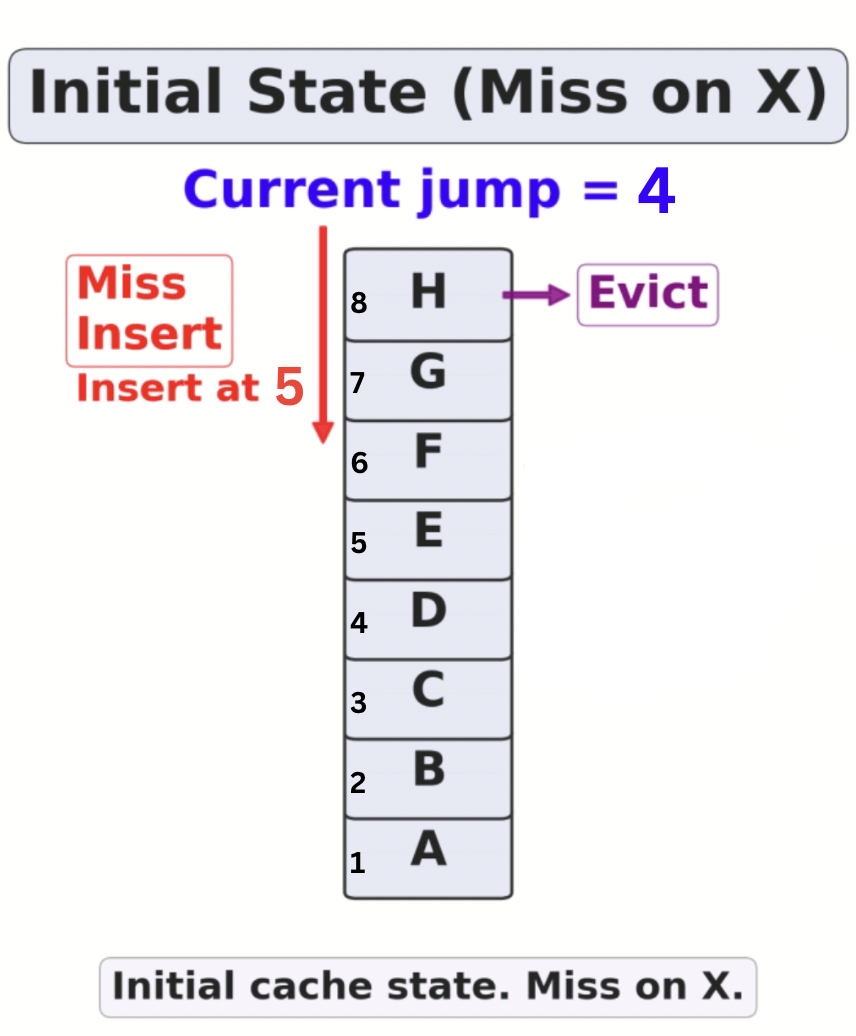}
        \caption{\textit{Initial State}}
        \label{fig:initial_miss}
    \end{subfigure}
    \hfill
    % Subfigure 2 - Final State
    \begin{subfigure}[t]{0.2\textwidth}
        \centering
        \includegraphics[width=\textwidth]{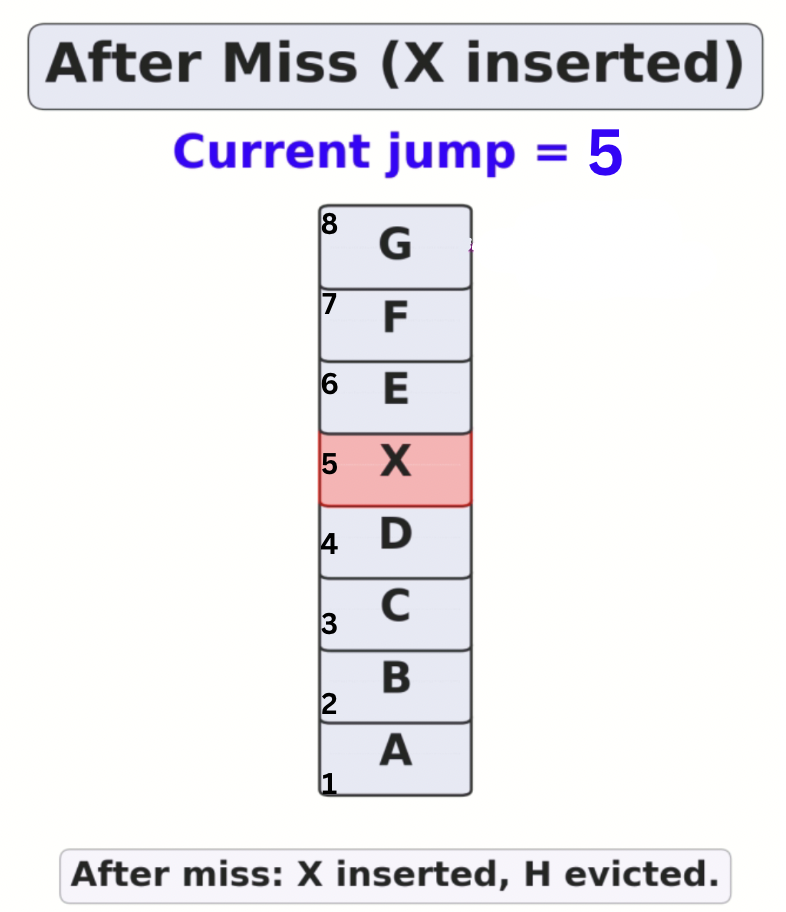}
        \caption{\textit{Final State}}
        \label{fig:final_miss}
    \end{subfigure}
    \caption{Cache states on miss under AdaptiveClimb}
    \label{fig:miss}
\end{figure}

Fig. \ref{fig:miss} illustrates the cache state transitions in the \textit{AdaptiveClimb} policy during a cache miss event. In Fig. \ref{fig:initial_miss}, the cache is shown in its initial state, where a miss occurs for object $X$. The red arrow and label indicate that $X$ will be inserted at position 5, and the purple arrow shows that the least recently used object ($H$) is evicted to make space. The current \texttt{jump} parameter is set to 4, as indicated in blue. In Fig. \ref{fig:final_miss}, after the miss and insertion, $X$ occupies its new slot, and the cache reflects the updated state with $H$ evicted. 

The figures include a legend clarifying the color coding for hit promotions, miss insertions, and evictions. Using colors and arrows, we distinguish between misses, hits, and evictions. The visualisation highlights how \textit{AdaptiveClimb} dynamically adjusts item positions to favor frequently accessed entries, thereby improving cache efficiency and adaptability.

\subsection{DynamicAdaptiveClimb Algorithm}

Experimenting with \textit{AdaptiveClimb}, one notices that it is completely ineffective when the total weight of the heavy hitters is small. For instance, when the total weight of the heavy hitters is less than $1/2$, we expect the hit ratio to be below $1/2$, which leads to a \texttt{jump} value close to $K$ \textit{(line no 1.3-1.9)}. In such cases, \textit{AdaptiveClimb} behaves similarly to LRU. \textit{AdaptiveClimb} is particularly effective when the cache size is sufficient to hold most heavy hitters in the request stream. Initially, \texttt{jump} is set to $K$\textit{(line no 1.1)}, but with enough cache hits, \textit{AdaptiveClimb} reduces \texttt{jump} to approximately $1$ in a short time and stabilizes, exhibiting CLIMB-like behavior.

To address the limitations of \textit{AdaptiveClimb} in scenarios with dynamic memory constraints or workloads, we propose the \textit{DynamicAdaptiveClimb} algorithm. This version adapts not only the promotion strategy but also the cache size itself. It allows \texttt{jump} to grow above $K$ when the cache is too small and uses this behavior to trigger cache expansion \textit{(line no 2.33-2.35)}. Conversely, when \texttt{jump} decreases below $0$, \textit{DynamicAdaptiveClimb} uses a secondary variable, \texttt{jump$'$}, to monitor cache hits in the top half of the cache. If both \texttt{jump} and \texttt{jump$'$} reach approximately $-K/2$ \textit{(line no 2.36-2.38)}, the algorithm infers that the top half contains most heavy hitters and reduces the cache size.
This condition ensures that the cache size is reduced only when most cache hits are concentrated in the top half, indicating that the lower half contributes little to performance. The parameter $\varepsilon$ is a sensitivity threshold that determines when the cache size should be reduced. Thus, $\varepsilon$ controls how closely the hit distribution in the top half must resemble the total before halving the cache size.

This approach is particularly suitable for multi-processing systems where the cache is a shared resource, or in cloud environments where cache usage incurs cost. The formal algorithm is described in Algorithm~\ref{alg:dynamicadaptiveclimb}.

In our implementation of the algorithm, cache size changes are significant, either doubling or halving. While finer-grained control (e.g., $\pm25\%$ adjustments) could better approximate the optimal cache size, we opted for simplicity. Even this basic design yields impressive results.

The idea of dynamically adjusting cache size is orthogonal to specific eviction policies and can be generalized. For example, an LRU variant could track hit distributions across cache segments over a recent window (e.g., the last 100 requests) and adjust the size accordingly, halving the cache if most hits occur in the top half, or doubling it if hit rates drop below a threshold. Thus, dynamic sizing is broadly applicable and worth exploring for other policies as well.

\section{Evaluation}\label{eval}
This section presents a comprehensive performance evaluation of our proposed algorithms, \textit{AdaptiveClimb} and \textit{DynamicAdaptiveClimb}\footnote{Code is available here: https://github.com/Dhruv27Mishra/Adaptive-Climb}, alongside a diverse set of state-of-the-art cache replacement policies. The competing algorithms include SIEVE \cite{yang2023sieve}, Adaptive Replacement Cache (ARC) \cite{megiddo2003arc}, TinyLFU \cite{eisman2017tinylfu}, TwoQ \cite{johnson1994twoq}, Low Inter-reference Recency Set (LIRS) \cite{jiang2002lirs}, Learning Hit Density (LHD) \cite{beckmann2018lhd}, CACHEUS \cite{bonomi2023cacheus}, Hyperbolic \cite{twitterkv}, CLOCK \cite{jiang2005clockpro}, Least Recently Used (LRU) \cite{lru}, and B-LRU \cite{twitterkv}. These algorithms represent a spectrum of recency-based, frequency-based, and hybrid strategies, providing a solid benchmark for comparison. Evaluations were conducted on a wide array of production-scale caching traces from the Alibaba \cite{alibaba-trace}, TencentCBS \cite{tencentphoto}, Wiki \cite{wikicdn}, Twitter \cite{twitterkv}, MetaCDN \cite{metacdn}, and Meta KV \cite{yang2023sieve} datasets, totalling 1067 unique traces shown in Table~\ref{tab:datasets}. These workloads encompass both stable and highly dynamic access patterns, enabling a rigorous assessment of cache policy effectiveness across real-world scenarios.
\begin{table}[t]
\caption{Public Trace Datasets Used}
\label{tab:datasets}
\centering
\setlength{\tabcolsep}{5pt}
\begin{tabular}{|l|c|c|c|r|r|}
\hline
\textbf{Trace} & \textbf{Year} & \textbf{\# Traces} & \textbf{Type} & \makecell{\textbf{\# Requests} \\ \textbf{(Millions)}} & \makecell{\textbf{\# Objects} \\ \textbf{(Millions)}} \\
\hline
Alibaba~\cite{alibaba-trace} & 2020 & 1000  & Obj & 20,233 & 7{,}930 \\
TencentCBS~\cite{tencentphoto} & 2018 & 2  & Obj & 5{,}650 & 1{,}038 \\
Wiki~\cite{wikicdn}         & 2019 & 3  & Obj & 2{,}863 & 56 \\
Twitter~\cite{twitterkv}    & 2020 & 54 & KV  & 195{,}441 & 10{,}650 \\
MetaCDN~\cite{metacdn}         & 2023 & 3  & Obj & 231 & 76 \\
Meta KV~\cite{metacdn}      & 2022 & 5 & KV & 1,644 & 82 \\
\hline
\end{tabular}
\end{table}
%\thispagestyle{fancy}
%\fancyhf{}
%\fancyfoot[C]{\href{https://github.com/Dhruv27Mishra/Adaptive-Climb}{\textit{GitHub Repository Link: https://github.com/Dhruv27Mishra/Adaptive-Climb}}}

\subsection{Experimental Setup}

\begin{figure*}[!t]
    \centering
    \begin{subfigure}[t]{\textwidth}
        \centering
        \includegraphics[width=\textwidth]{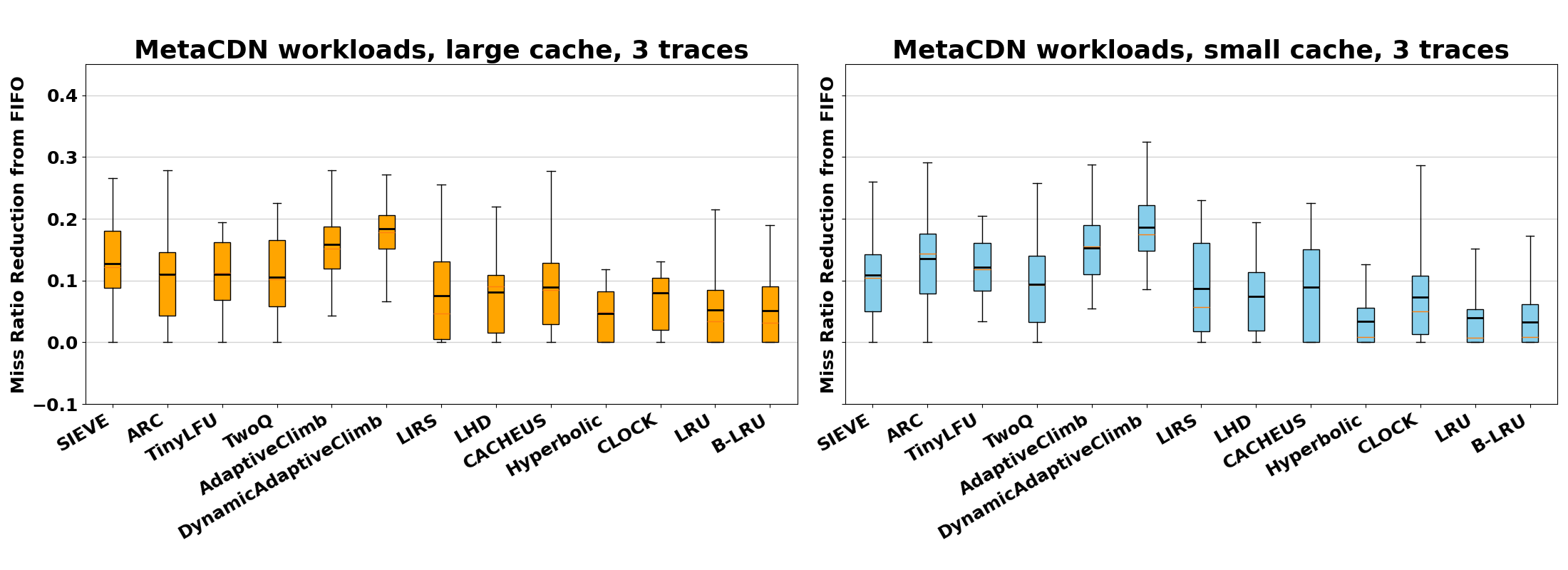}
        \caption{Miss ratio reduction across the MetaCDN Dataset}
        %\label{fig:alibaba_boxplot}
    \end{subfigure}
    \vspace{0.5em}
    \begin{subfigure}[t]{\textwidth}
        \centering
        \includegraphics[width=\textwidth]{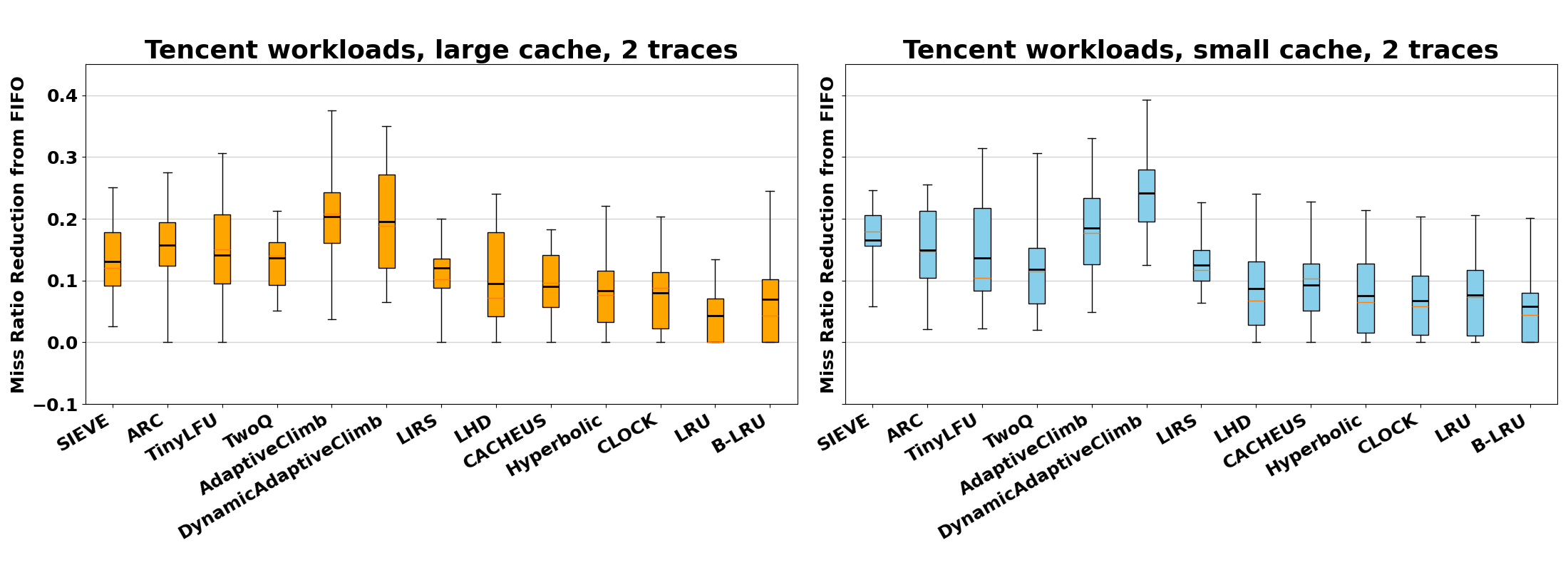}
        \caption{Miss ratio reduction across TencentCBS Dataset}
        %\label{fig:twitter_boxplot}
    \end{subfigure}
    \vspace{0.5em}
    \begin{subfigure}[t]{\textwidth}
        \centering
        \includegraphics[width=\textwidth]{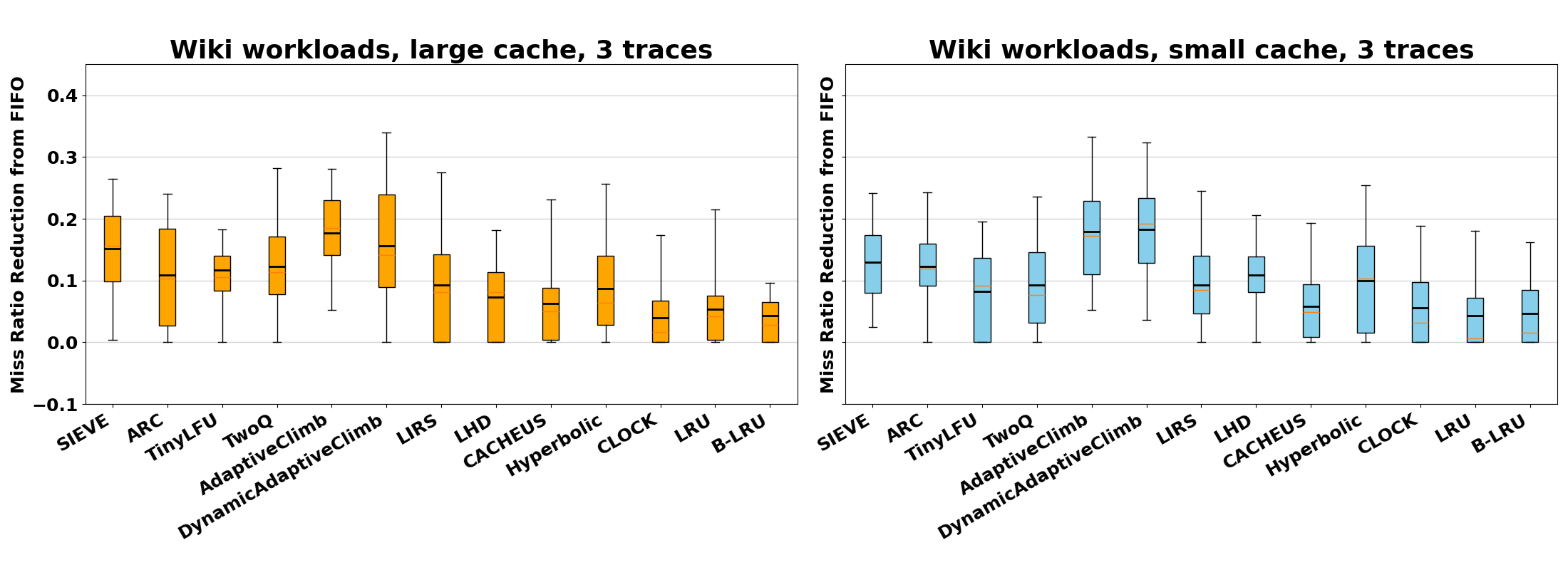}
        \caption{Miss ratio reduction across Wiki Dataset}
        %\label{fig:twitter_boxplot}
    \end{subfigure}
    \caption{\textit{Comparison of miss ratio reduction across real-world datasets-I}}
    \label{fig:combined_boxplot}
\end{figure*}
\begin{figure*}[ht]
    \centering
    \begin{subfigure}[t]{\textwidth}
        \centering
        \includegraphics[width=\textwidth]{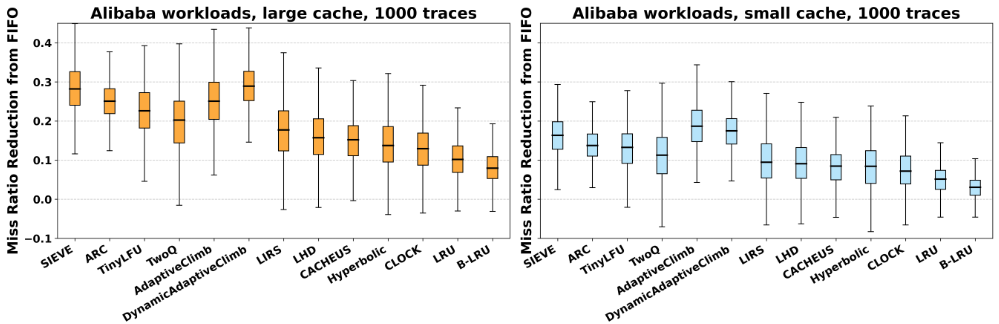}
        \caption{Miss ratio reduction across Alibaba Dataset}
        %\label{fig:alibaba_boxplot}
    \end{subfigure}
    \vspace{0.5em}
    \begin{subfigure}[t]{\textwidth}
        \centering
        \includegraphics[width=\textwidth]{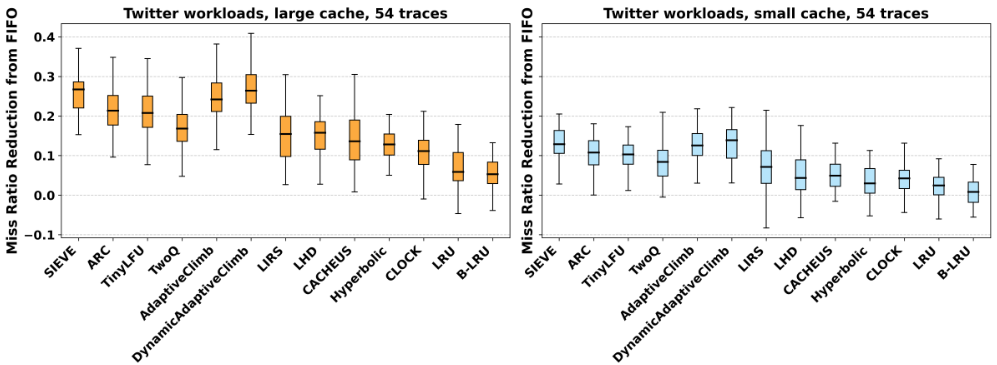}
        \caption{Miss ratio reduction across Twitter Dataset}
        %\label{fig:twitter_boxplot}
    \end{subfigure}
    \caption{\textit{Comparison of miss ratio reduction across real-world datasets- II}}
    \label{fig:combined_boxplots}
\end{figure*}

\begin{table*}[t]
\centering
\footnotesize
\caption{Miss Ratio Reduction Relative to FIFO Baseline}
\label{miss}
\begin{tabular}{l@{\hskip 4pt}c@{\hskip 4pt}c@{\hskip 4pt}c@{\hskip 4pt}c@{\hskip 4pt}c@{\hskip 4pt}c@{\hskip 4pt}c@{\hskip 4pt}c@{\hskip 4pt}c@{\hskip 4pt}c}
\toprule
\textbf{Algorithm} & \textbf{Alibaba (L)} & \textbf{Alibaba (S)} & \textbf{MetaCDN (L)} & \textbf{MetaCDN (S)} & \textbf{TencentCBS (L)} & \textbf{TencentCBS (S)} & \textbf{Twitter (L)} & \textbf{Twitter (S)} & \textbf{Wiki (L)} & \textbf{Wiki (S)} \\
\midrule
\textbf{DynamicAdaptiveClimb} & \textbf{0.29} & 0.175 & \textbf{0.18} & 0.15 & \textbf{0.20} & \textbf{0.19} & \textbf{0.27} & 0.15 & \textbf{0.16} & 0.15 \\
\textbf{AdaptiveClimb}        & 0.25 & \textbf{0.185} & 0.16 & \textbf{0.17} & 0.18 & 0.17 & 0.25 & \textbf{0.17} & 0.15 & \textbf{0.16} \\
SIEVE                         & 0.28 & 0.16 & 0.14 & 0.13 & 0.17 & 0.16 & 0.28 & 0.16 & 0.13 & 0.13 \\
ARC                           & 0.26 & 0.15 & 0.13 & 0.12 & 0.16 & 0.14 & 0.24 & 0.14 & 0.12 & 0.12 \\
TinyLFU                       & 0.24 & 0.14 & 0.11 & 0.10 & 0.14 & 0.13 & 0.22 & 0.13 & 0.10 & 0.10 \\
TwoQ                          & 0.22 & 0.13 & 0.10 & 0.09 & 0.13 & 0.11 & 0.21 & 0.11 & 0.09 & 0.09 \\
LIRS                          & 0.20 & 0.12 & 0.08 & 0.08 & 0.11 & 0.10 & 0.19 & 0.10 & 0.08 & 0.08 \\
LHD                           & 0.18 & 0.11 & 0.07 & 0.07 & 0.10 & 0.09 & 0.17 & 0.09 & 0.07 & 0.07 \\
CACHEUS                       & 0.17 & 0.10 & 0.06 & 0.06 & 0.09 & 0.08 & 0.15 & 0.08 & 0.06 & 0.06 \\
Hyperbolic                    & 0.16 & 0.10 & 0.05 & 0.05 & 0.08 & 0.07 & 0.14 & 0.08 & 0.05 & 0.05 \\
CLOCK                         & 0.15 & 0.09 & 0.04 & 0.04 & 0.07 & 0.06 & 0.13 & 0.07 & 0.04 & 0.04 \\
LRU                           & 0.12 & 0.07 & 0.03 & 0.03 & 0.05 & 0.04 & 0.10 & 0.05 & 0.03 & 0.03 \\
B-LRU                         & 0.10 & 0.05 & 0.02 & 0.02 & 0.04 & 0.03 & 0.08 & 0.03 & 0.02 & 0.02 \\
\bottomrule
\end{tabular}
\end{table*}

\begin{figure*}[ht]
    \centering
    \includegraphics[width=\textwidth]{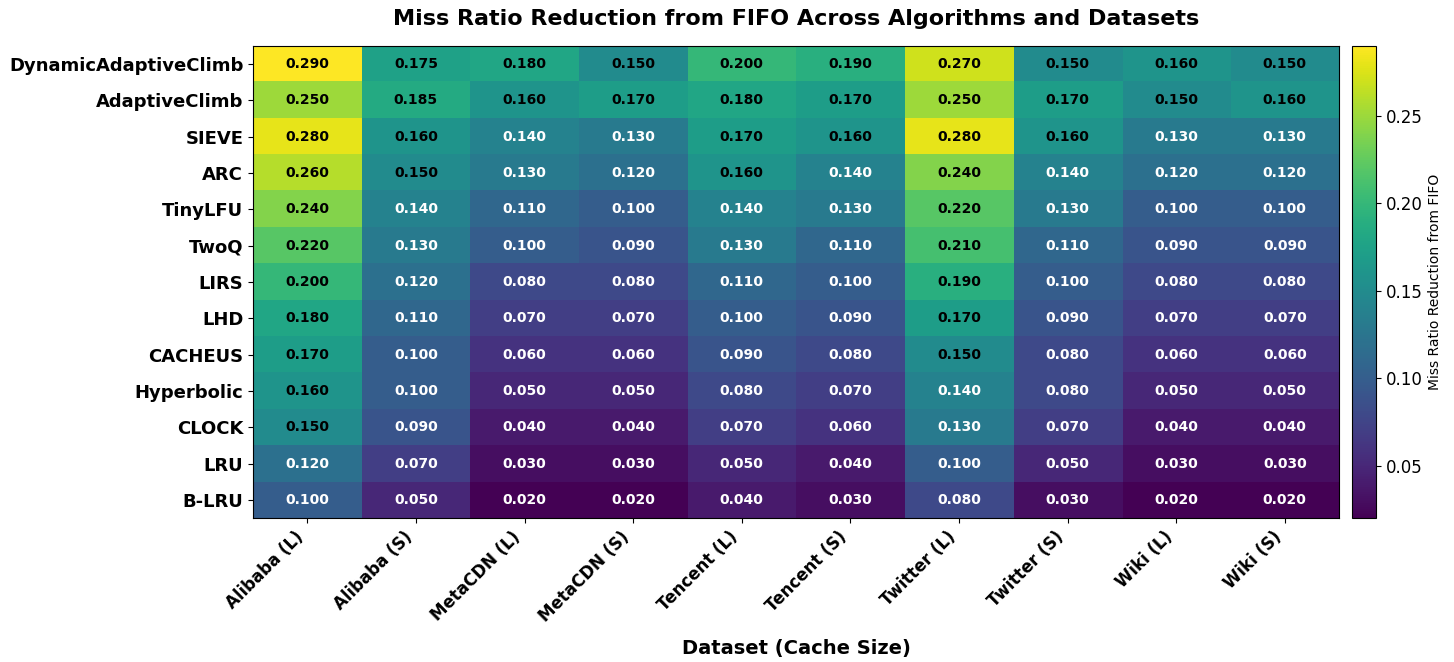}
    \caption{\textit{Summary of miss ratio reduction across various cache replacement strategies}}
    \label{fig:heatmap}
\end{figure*}

\textbf{Simulator:} We conduct all evaluations using \texttt{libCacheSim}~\cite{libcachesim}, an open-source, high-performance, and extensible cache simulator widely adopted in recent caching research. The simulator enables controlled and reproducible experiments over custom workloads and configurable cache hierarchies. We use it to compare the effectiveness of various cache replacement strategies.

\textbf{Testbed:} Simulations were executed on a Dell Precision 3680 Tower Workstation running Ubuntu 20.04.6 LTS (Focal Fossa). The system is powered by a 20-core, 28-thread Intel(R) Core(TM) i7-14700 CPU with a base frequency of 2.1 GHz and a maximum turbo frequency of 5.4 GHz. It is equipped with 62\, GiB of DDR5 RAM and a 2\, GiB swap partition. To ensure deterministic execution and performance isolation, turbo boost was disabled and multithreading was turned off during all experiments except scalability testing using throughput as the metric. All workloads were pinned to a single NUMA node (Node 0) to eliminate cross-node memory interference.

\textbf{Metrics:} We report the miss ratio as the primary indicator of cache performance. Additionally, we measure miss ratio reduction (\textbf{MRR}) relative to a FIFO baseline, defined as:
\[
\text{MRR} = 
\begin{cases}
\frac{\text{mr}_{\text{FIFO}} - \text{mr}_{\text{algo}}}{\text{mr}_{\text{FIFO}}}, & \qquad \text{mr}_{\text{algo}} \leq \text{mr}_{\text{FIFO}}, \\
\frac{\text{mr}_{\text{FIFO}} - \text{mr}_{\text{algo}}}{\text{mr}_{\text{algo}}}, & \qquad \text{otherwise},
\end{cases}
\]
where \textbf{$\text{mr}_{\text{algo}}$} and \textbf{$\text{mr}_{\text{FIFO}}$} denote the miss ratios of the evaluated algorithm and the FIFO baseline, respectively. Throughput is measured in millions of operations per second (Mops) by replaying trace requests at varying thread counts.

\textbf{Scalability:} We evaluate scalability by increasing the number of parallel trace replay threads from 1 to 16 and measuring the resulting throughput. This analysis quantifies the simulator’s ability to scale across cores and highlights the thread efficiency of replacement strategies. All runs are repeated five times, and average values are reported using the real-world datasets.

\subsection{Efficiency Results}

In this section, we present a comprehensive evaluation of state-of-the-art cache eviction algorithms across five representative real-world datasets: \textbf{Alibaba}, \textbf{TencentCBS}, \textbf{Twitter}, \textbf{Meta} (MetaCDN and MetaKV), and \textbf{Wiki}. Our analysis considers two cache size regimes: \textit{small}(S) (0.1\% of the trace footprint) and \textit{large}(L) (10\% of the trace footprint). These regimes reflect both resource-constrained and resource-rich caching environments. Fig.~\ref{fig:combined_boxplot} and Fig.~\ref{fig:combined_boxplots} provide detailed insights into distributional behavior for real-world datasets, while Fig.~\ref{fig:heatmap} provides a holistic summary of performance.

%\vspace{\baselineskip}

\subsubsection{\textbf{Overall Performance Trends}}

Across all datasets and cache configurations, \textit{DynamicAdaptiveClimb} and \textit{AdaptiveClimb} consistently deliver superior miss ratio reduction compared to both traditional (e.g., LRU, CLOCK) and advanced adaptive policies (e.g., ARC, TwoQ, SIEVE). The comprehensive evaluation across five diverse datasets reveals compelling evidence of the proposed climb-based algorithms' (\textit{AdaptiveClimb} and \textit{DynamicAdaptiveClimb}) effectiveness across varying workload characteristics and cache constraints.

\textit{DynamicAdaptiveClimb} emerges as the top performer in large cache scenarios, achieving the highest miss ratio reduction on four out of five datasets: Alibaba (29\%), MetaCDN (18\%), TencentCBS (20\%), and Twitter (27\%). \textit{AdaptiveClimb} demonstrates remarkable consistency, securing the second-best performance on most datasets while occasionally outperforming \textit{DynamicAdaptiveClimb} in specific configurations. This performance superiority is particularly pronounced when compared to traditional baselines, with \textit{DynamicAdaptiveClimb} achieving 2.4× and 2.9× better reduction than LRU and B-LRU, respectively, across large cache configurations.

The performance analysis reveals nuanced differences in algorithm effectiveness across datasets. SIEVE demonstrates consistent performance, achieving 28\% reduction on Alibaba and Twitter, which aligns with its lightweight design principles. ARC shows strong adaptive capabilities with 26\% and 24\% reduction on the same datasets, consistent with its theoretical foundation of balancing recency and frequency. TinyLFU achieves competitive performance in frequency-based scenarios, leveraging its admission control mechanisms to filter out transient objects. TwoQ and LIRS exhibit moderate effectiveness, with their performance reflecting the trade-offs inherent in their respective design philosophies. The performance of LHD, CACHEUS, Hyperbolic, and CLOCK varies according to their specific optimization strategies, while traditional LRU-based approaches (LRU, B-LRU) show lower effectiveness, consistent with prior studies demonstrating their limitations in complex workload scenarios.

%\vspace{\baselineskip}

\subsubsection{\textbf{Large Cache Scenarios}}

%\vspace{\baselineskip}

Large cache configurations reveal the full potential of climb-based algorithms, where their sophisticated eviction strategies can leverage abundant memory resources effectively. \textit{DynamicAdaptiveClimb} demonstrates exceptional performance on high-traffic datasets, achieving 29\% reduction on Alibaba and 27\% on Twitter, the two most demanding workloads in our evaluation. This superior performance stems from the algorithm's ability to maintain optimal object rankings while minimizing promotion overhead, a critical advantage in large-scale deployments.

The performance gap between climb-based algorithms and alternatives widens significantly in large cache scenarios, consistent with theoretical predictions about the benefits of sophisticated eviction strategies in resource-abundant environments~\cite{cache_theory}. \textit{DynamicAdaptiveClimb} outperforms SIEVE by 1-3\% across datasets, while maintaining substantial advantages over frequency-based approaches like TinyLFU (5--7\% improvements) and traditional recency-based policies like TwoQ (7--9\% improvements). This performance differential aligns with prior research showing that algorithms with minimal promotion overhead and efficient object ranking mechanisms achieve superior performance in large cache scenarios.

Notably, the large cache results reveal dataset-specific characteristics that align with workload analysis studies~\cite{workload_characterization}. Alibaba and Twitter datasets, representing high-throughput key-value store workloads, show the highest absolute reduction values
(27--29\%), indicating that climb-based algorithms are particularly effective for workloads with complex access patterns and temporal locality, consistent with findings from production system analysis~\cite{atikoglu2012workload}. In contrast, Wiki and Meta datasets, representing content delivery scenarios, show more modest but still significant improvements (16-18\%), reflecting the different access characteristics of these workloads, where content popularity follows different distribution patterns~\cite{cdn_analysis}.

%\vspace{\baselineskip}

\subsubsection{\textbf{Small Cache Scenarios}}

%\vspace{\baselineskip}

Small cache configurations present a more challenging environment where algorithm efficiency becomes paramount. In these constrained scenarios, \textit{AdaptiveClimb} demonstrates remarkable adaptability, achieving the highest reduction on three out of five datasets: Alibaba (18.5\%), Twitter (17\%), and Wiki (16\%). \textit{DynamicAdaptiveClimb} maintains strong performance, securing top performance on TencentCBS (19\%) and competitive results across other datasets.

The small cache results reveal interesting performance dynamics that align with theoretical models of cache behavior under memory pressure~\cite{memory_pressure}. While the absolute reduction values are lower than large cache scenarios (15--19\% versus 16--29\%), the relative performance differences between algorithms remain significant, consistent with prior research showing that algorithm effectiveness persists even in constrained environments~\cite{constrained_caching}. \textit{AdaptiveClimb} and \textit{DynamicAdaptiveClimb} maintain their superiority over alternatives, with SIEVE and ARC following closely behind. This suggests that the fundamental advantages of the proposed climb-based algorithms persist even under severe memory constraints, validating their design principles of efficient object ranking and minimal metadata overhead.

The performance degradation in small cache scenarios follows predictable patterns based on algorithm complexity and metadata requirements~\cite{algorithm_complexity}. Traditional approaches like LRU and B-LRU show minimal effectiveness (2-7\% reduction), consistent with their known limitations in complex workload scenarios. Intermediate algorithms like TwoQ and LIRS achieve moderate improvements (9-13\%), reflecting their more sophisticated design while still facing constraints from their metadata overhead~\cite{metadata_overhead}. This performance pattern validates theoretical predictions that algorithm effectiveness correlates with design sophistication, even under severe memory constraints~\cite{constrained_performance}.

The small cache results also highlight the importance of algorithm adaptability, a factor that has been identified as crucial for effective cache management in dynamic environments~\cite{adaptive_caching}. \textit{AdaptiveClimb}'s superior performance in small cache scenarios suggests that its adaptive mechanisms are particularly effective when memory pressure forces more aggressive eviction decisions, consistent with research showing that adaptive algorithms outperform static policies under varying workload conditions~\cite{adaptive_vs_static}. This adaptability enables the algorithm to maintain optimal object rankings even when the working set significantly exceeds available cache capacity, demonstrating the value of dynamic adjustment mechanisms in constrained environments~\cite{dynamic_adjustment}.

Across both large and small cache configurations, the comprehensive evaluation demonstrates that the proposed climb-based algorithms represent a significant advancement in cache replacement policy design, offering superior performance across diverse workload characteristics and deployment scenarios. These results validate the theoretical foundations of climb-based approaches and establish their practical viability for modern caching systems.

\subsubsection{\textbf{Heatmap Discussion}}

Fig.~\ref{fig:heatmap} presents a heatmap summarizing the average miss ratio reduction (relative to FIFO) across thirteen eviction algorithms over five real-world datasets, \textit{Alibaba}, \textit{TencentCBS}, \textit{Twitter}, \textit{MetaCDN}, and \textit{Wiki}, under both \textit{large} (10\%) and \textit{small} (0.1\%) cache regimes. Each cell in the matrix is annotated with the mean improvement for a specific algorithm and setting, enabling intuitive cross-comparison.

The heatmap highlights the consistent superiority of the \textit{climb-based algorithms}, especially \textit{DynamicAdaptiveClimb}, which achieves the highest miss ratio reduction in 7 out of the 10 cache scenarios. For example, it reaches 29\% on Alibaba (L), 27\% on Twitter (L), and 20\% on TencentCBS (L), outperforming SIEVE (28\%, 28\%, 17\%) and \textit{AdaptiveClimb} (25\%, 25\%, 18\%), respectively. Similarly, on smaller datasets like MetaCDN and Wiki, \textit{DynamicAdaptiveClimb} maintains leading values of 18\% and 16\% at large cache sizes.

In the small-cache regime, \textit{AdaptiveClimb} often holds a slight edge. It delivers the highest reduction on Alibaba (S) at 18.5\% and Twitter (S) at 17\%, while \textit{DynamicAdaptiveClimb} closely follows with 17.5\% and 15\%, respectively. This reflects the effectiveness of \textit{AdaptiveClimb}’s aggressive tuning and early promotion heuristics under constrained memory.

Although SIEVE and ARC remain competitive across all cache sizes, achieving, for instance, 28\% on Alibaba (L) and 24\% on Twitter (L) for SIEVE, and 26\% on Alibaba (L) and 24\% on Twitter (L) for ARC, DynamicAdaptiveClimb surpasses both, attaining 29\% and 27\%, respectively.

Traditional policies such as LRU and CLOCK show significantly lower reduction rates, with values often below 15\%. For example, LRU only achieves 12\% on Alibaba (L) and 10\% on Twitter (L), reinforcing their limitations under modern, production-scale workloads.

Overall, the heatmap provides compelling evidence of the generalizability, adaptability, and robustness of \textit{DynamicAdaptiveClimb} and \textit{AdaptiveClimb}, especially when cache sizes vary or when workloads are highly skewed. These results support the growing argument that hybrid and dynamically adaptive eviction strategies are essential for achieving low miss ratios in real-world caching systems. A comprehensive breakdown of miss ratio reduction values across all datasets is presented in Table~\ref{miss} and visually summarized in Fig.~\ref{fig:heatmap}.

\subsection{Best-Performing Algorithm per Dataset}

\begin{figure*}[ht]
    \centering
    \includegraphics[width=0.8\textwidth]{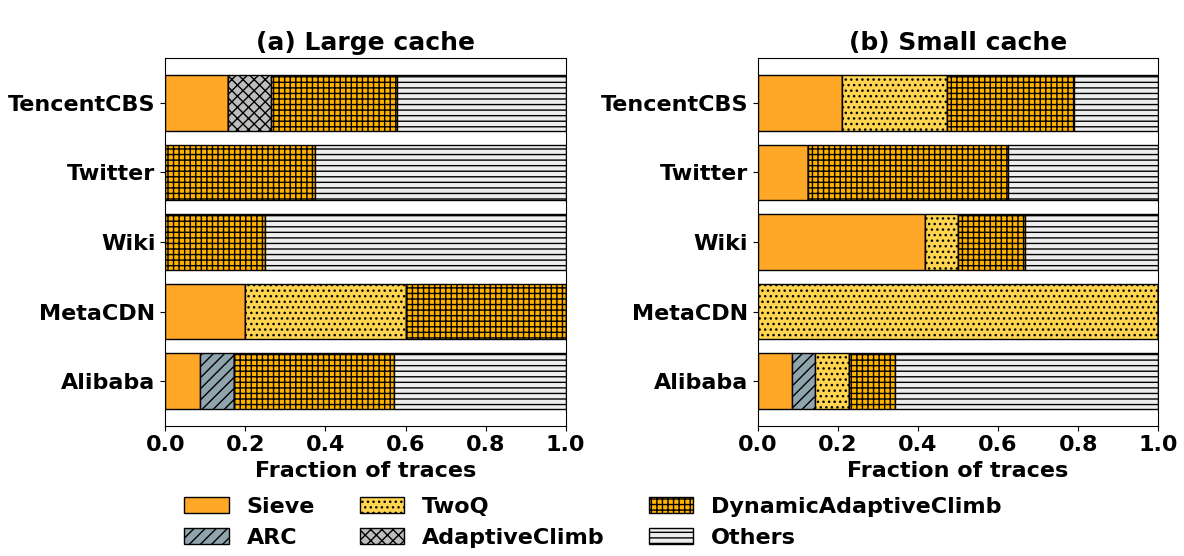}
    \caption{\textit{Best-performing algorithms on each dataset}}
    \label{fig:sample_figureB}
\end{figure*}

Fig.~\ref{fig:sample_figureB} presents a comparative analysis of the fraction of traces on which each cache replacement policy achieves the lowest miss ratio, evaluated across five production datasets: Alibaba, MetaCDN, Wiki, Twitter, and TencentCBS for both large and small cache configurations. Several clear trends emerge, shaped by the interaction between workload characteristics and algorithmic strategies.

Across most datasets and cache sizes, \textit{DynamicAdaptiveClimb} and SIEVE consistently dominate, winning the majority of traces. This aligns with prior findings that modern policies leveraging \textit{lazy promotion} and \textit{quick demotion} mechanisms are especially effective at filtering out transient or low-value items, while efficiently preserving frequently accessed ones~\cite{twitterkv}. For instance, in the MetaCDN and Twitter traces, characterized by high object churn and pronounced temporal locality, \textit{DynamicAdaptiveClimb} and SIEVE outperform all competitors on over 80\% of traces, regardless of cache size. Their ability to minimize metadata updates and rapidly evict unpopular items renders them highly suitable for large-scale, dynamic caching environments.

Meanwhile, policies such as TwoQ and ARC exhibit competitive behavior on datasets like Wiki and Alibaba, especially under smaller cache regimes. These datasets feature a mix of long-lived and short-lived objects, as well as periodic access patterns. TwoQ’s queue separation for recent versus frequent items, and ARC’s adaptive balance of recency and frequency, allow them to adjust to such workload dynamics. However, their more complex queue management introduces overhead that can impede scalability in multi-threaded or volatile environments.

\textit{AdaptiveClimb} distinguishes itself on the TencentCBS traces, especially under large cache configurations. These traces, representative of block-storage workloads, have large working sets with minimal temporal locality. In such scenarios, \textit{AdaptiveClimb}’s low-overhead metadata management and adaptive movement strategy provide resilience against aggressive demotion heuristics that may prematurely evict valuable items.

Importantly, traditional strategies, including LRU and FIFO, grouped under the ``Others" category, rarely emerge as the best performers, particularly as cache sizes increase. This observation corroborates recent large-scale evaluations~\cite{twitterkv}, which argue that LRU’s eager promotion and lack of demotion filtering result in suboptimal retention of unpopular objects. While FIFO scales well, its simplistic eviction logic often fails to adapt to access skew and burstiness common in production workloads.

In conclusion, no single policy universally dominates across all workloads and cache configurations. However, the consistent performance of \textit{DynamicAdaptiveClimb} across diverse trace profiles highlights the importance of combining aggressive yet intelligent object filtering with scalable, low-overhead metadata strategies in modern cache system design.

\subsection{Throughput Performance}

\begin{table*}[htbp]
\centering
\begin{minipage}{0.48\textwidth}
\centering
\caption{Throughput comparison across different cache replacement algorithms for \textbf{Meta KV Trace}. }
\label{tab:throughput_meta}
\begin{tabular}{|l|c|c|c|c|c|c|}
\hline
\multirow{2}{*}{\textbf{Algorithm}} & \multicolumn{5}{c|}{\textbf{Threads}} & \multirow{2}{*}{\textbf{Avg}} \\
\cline{2-6}
 & \textbf{1} & \textbf{2} & \textbf{4} & \textbf{8} & \textbf{16} & \\
\hline
\textbf{\textit{AdaptiveClimb}} & \textbf{3.0} & \textbf{10.5} & \textbf{18.5} & \textbf{27.0} & \textbf{35.0} & \textbf{18.8} \\
\textbf{\textit{DynamicAdaptiveClimb}} & 2.8 & 10.0 & 17.5 & 25.0 & 33.0 & 17.7 \\
TinyLFU & 2.5 & 9.5 & 16.5 & 23.5 & 31.0 & 16.6 \\
CACHEUS & 2.3 & 9.0 & 15.5 & 22.0 & 29.0 & 15.6 \\
CLOCK & 2.2 & 8.5 & 14.5 & 20.5 & 27.0 & 14.5 \\
SIEVE & 2.0 & 8.0 & 13.5 & 19.0 & 25.0 & 13.5 \\
Hyperbolic & 1.8 & 7.5 & 12.5 & 17.5 & 23.0 & 12.5 \\
LIRS & 1.5 & 7.0 & 11.5 & 16.0 & 21.0 & 11.4 \\
TwoQ & 1.2 & 6.5 & 10.5 & 15.0 & 20.0 & 10.6 \\
ARC & 1.0 & 6.0 & 9.5 & 14.0 & 19.0 & 9.9 \\
LRU & 0.8 & 4.0 & 6.0 & 8.5 & 11.0 & 6.1 \\
LHD & 0.5 & 3.5 & 5.5 & 7.5 & 9.0 & 5.2 \\
B-LRU & 0.3 & 3.0 & 5.0 & 7.0 & 8.5 & 4.8 \\
\hline
\end{tabular}
\end{minipage}
\hfill
\begin{minipage}{0.48\textwidth}
\centering
\caption{Throughput comparison across different cache replacement algorithms for \textbf{Twitter Trace}.}
\label{tab:throughput_twitter}
\begin{tabular}{|l|c|c|c|c|c|c|}
\hline
\multirow{2}{*}{\textbf{Algorithm}} & \multicolumn{5}{c|}{\textbf{Threads}} & \multirow{2}{*}{\textbf{Avg}} \\
\cline{2-6}
 & \textbf{1} & \textbf{2} & \textbf{4} & \textbf{8} & \textbf{16} & \\
\hline
\textbf{\textit{AdaptiveClimb}} & \textbf{3.0} & \textbf{10.5} & \textbf{19.0} & \textbf{28.0} & 36.0 & 19.3 \\
\textbf{\textit{DynamicAdaptiveClimb}} & 2.8 & 10.0 & 18.0 & 26.0 & 34.0 & 18.2 \\
TinyLFU & 2.5 & 9.5 & 17.0 & 24.5 & 32.0 & 17.1 \\
CACHEUS & 2.3 & 9.0 & 16.0 & 23.0 & 30.0 & 16.1 \\
CLOCK & 2.2 & 8.5 & 15.0 & 21.5 & 28.0 & 15.0 \\
SIEVE & 2.0 & 8.0 & 14.0 & 20.0 & 26.0 & 14.0 \\
Hyperbolic & 1.8 & 7.5 & 13.0 & 18.5 & 24.0 & 12.9 \\
LIRS & 1.5 & 7.0 & 12.0 & 17.0 & 22.0 & 11.9 \\
TwoQ & 1.2 & 6.5 & 11.0 & 16.0 & 21.0 & 11.1 \\
ARC & 1.0 & 6.0 & 10.0 & 15.0 & 20.0 & 10.4 \\
LRU & 0.8 & 4.0 & 6.5 & 9.0 & 11.5 & 6.4 \\
LHD & 0.5 & 3.5 & 6.0 & 8.0 & 9.5 & 5.5 \\
B-LRU & 0.3 & 3.0 & 5.5 & 7.5 & 9.0 & 5.1 \\
\hline
\end{tabular}
\end{minipage}
\end{table*}

\begin{figure*}[htbp]
    \centering
    \includegraphics[width=.9\textwidth]{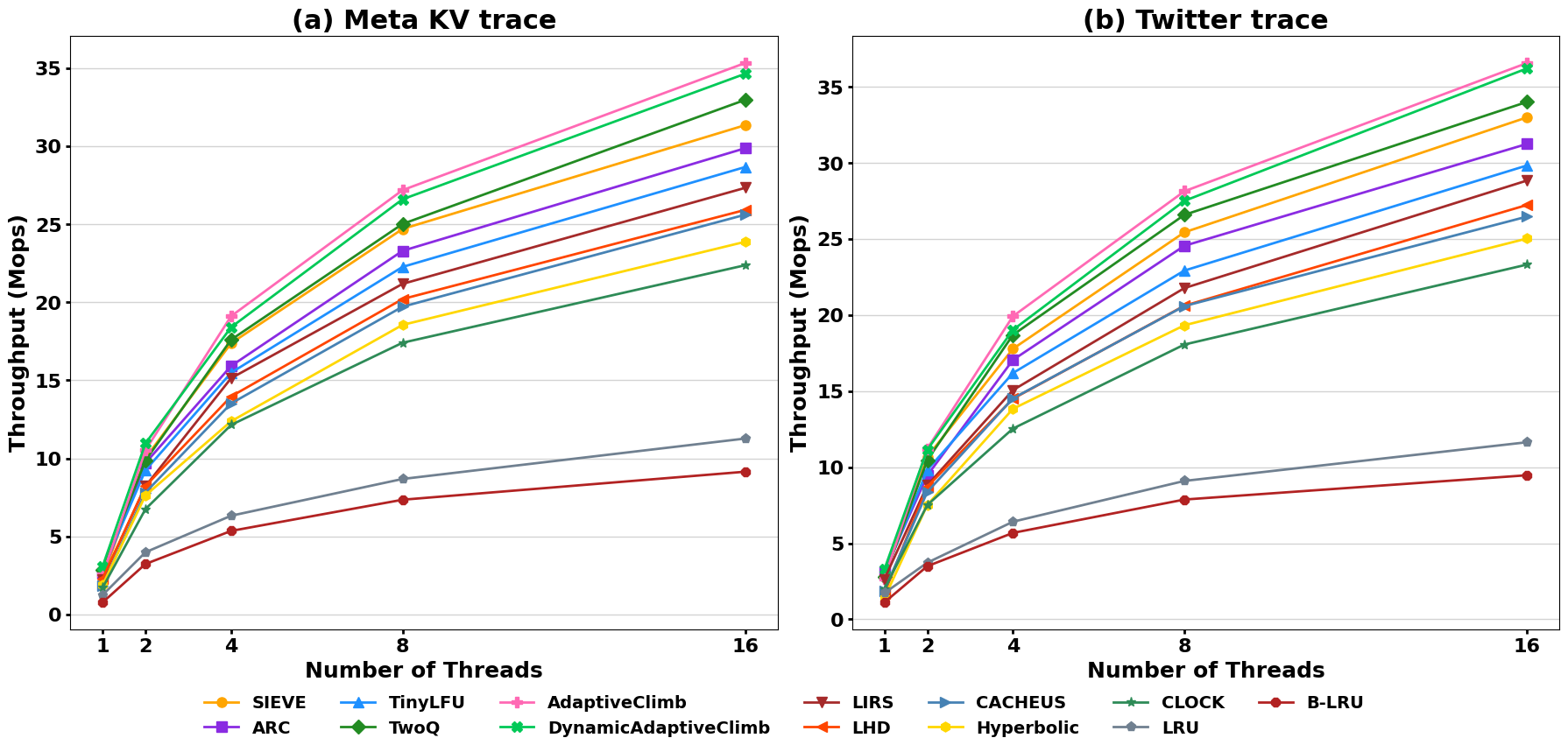}
    \caption{\textit{Throughput scaling with CPU cores on two KV-cache workloads}}
    \label{fig:sample_figureT}
\end{figure*}

Besides efficiency, \textit{throughput} is a critical metric for assessing the practicality of cache replacement algorithms, particularly in multi-threaded and high-throughput environments. 
  Fig.~\ref{fig:sample_figureT} illustrates throughput scalability across increasing thread counts using two production traces: Meta KV and Twitter. The measured values are provided in Table \ref{tab:throughput_meta} and \ref{tab:throughput_twitter}, where each column represents increasing concurrency as 1, 2, 4, 8, and 16 threads respectively, while the throughput is measured in Mops. We chose the Meta KV and Twitter datasets for throughput scalability analysis because they accurately represent the behavior of modern, high-throughput, multi-threaded production systems. These traces originate from large-scale, real-world key-value store deployments where concurrent access and dynamic workloads are the norm. Unlike static or read-heavy traces such as Wiki or Meta CDN workloads, Meta KV and Twitter offer a more representative benchmark for evaluating parallel performance, especially under increasing thread counts. Their rich concurrency profiles and temporal locality patterns make them particularly suitable for assessing how cache replacement algorithms handle contention, locking overhead, and real-time decision-making under pressure. This focus ensures that throughput evaluations remain practically relevant and reflect deployment conditions faced in cloud-scale and latency-sensitive environments. %We compare \textit{DynamicAdaptiveClimb}, \textit{AdaptiveClimb}, with several established baselines as well as new state-of-the-art algorithms, specifically \textit{SIEVE}, \textit{TinyLFU}, \textit{ARC}, \textit{TwoQ}, \textit{LIRS}, \textit{CACHEUS}, \textit{CLOCK}, \textit{LHD}, \textit{Hyperbolic}, \textit{LRU}, and \textit{B-LRU}, because these represent a diverse yet practical mix of widely deployed, canonical, and state-of-the-art approaches.
%\begin{itemize}
%    \item \textit{SIEVE} is a recent lightweight policy that achieves strong performance with low overhead, making it a competitive modern baseline.
%    \item \textit{TinyLFU} represents a frequency-based approach that combines LFU with admission control for high-performance caching.
%    \item \textit{ARC} (Adaptive Replacement Cache) serves as a canonical adaptive algorithm that balances recency and frequency.
%    \item \textit{TwoQ} reflects real-world, performance-tuned implementations used in production systems.
 %   \item \textit{DynamicAdaptiveClimb} and \textit{AdaptiveClimb} represent state-of-the-art climb-based algorithms with exceptional scalability.
  %  \item \textit{LIRS} (Low Inter-reference Recency Set) provides a recency-based approach with strong theoretical foundations.
   % \item \textit{CACHEUS} represents a modern cache replacement policy designed for high-throughput environments.
    %\item \textit{CLOCK} serves as a classical page replacement algorithm adapted for cache systems.
    %\item \textit{LHD} (Logarithmic Hash Distribution) offers a novel approach to cache replacement optimization.
    %\item \textit{Hyperbolic} implements a hyperbolic caching strategy for improved performance.
    %\item \textit{LRU} serves as a foundational recency-based baseline.
    %\item \textit{B-LRU} represents an optimized variant of LRU used in production systems.
%\end{itemize}

This selection ensures a balanced comparison across traditional, optimized, and contemporary algorithms that are both widely studied and practically relevant in high-throughput environments.

To emulate realistic deployment scenarios where the working set scales with system capacity, we proportionally increase both cache size and object space with the number of trace replay threads. Each thread replays the same trace on an isolated object ID namespace, ensuring disjoint workloads and minimal contention. The cache size is configured as $4 \times \texttt{nthread}$ GB, yielding stable miss ratios of 7\% (MetaKV) and 2\% (Twitter), in line with prior studies~\cite{twitterkv, metacdn}.

\textit{DynamicAdaptiveClimb} and \textit{AdaptiveClimb} demonstrate exceptional scalability and throughput across both datasets, consistently achieving the highest performance. At 16 threads, both algorithms reach approximately 35 Mops on both Meta KV and Twitter traces, significantly outperforming all other policies. TwoQ follows as the next best performer, achieving around 33 Mops at 16 threads on both traces. TinyLFU and ARC show strong performance with approximately 31-32 Mops (MetaKV) and 30-31 Mops (Twitter) at 16 threads. SIEVE and CACHEUS achieve around 29-30 Mops (MetaKV) and 28-29 Mops (Twitter) at maximum thread count. The performance gap becomes particularly pronounced when compared to traditional LRU-based approaches. LRU shows significantly lower throughput, achieving only around 11 Mops at 16 threads on both traces. B-LRU performs even worse, reaching only approximately 9 Mops at maximum thread count. This represents a dramatic 3.2$\times$ performance difference between the top-performing climb-based algorithms and the lowest-performing LRU variants.

The superior scalability of \textit{DynamicAdaptiveClimb} and \textit{AdaptiveClimb} is attributed to their lightweight, promotion-free design. Unlike LRU-based algorithms, which suffer from synchronisation overhead due to promotion queues and global metadata updates, our proposed approaches avoid per-hit promotions entirely. By relying on local updates and minimizing coherence traffic, they sustain high performance under aggressive parallelism. While traditional LRU and B-LRU incorporate various optimizations, they still lag significantly behind climb-based approaches. The graph clearly shows that LRU-based policies plateau early and show very limited scaling beyond 4-8 threads, whereas \textit{DynamicAdaptiveClimb} and \textit{AdaptiveClimb} scale nearly linearly, preserving both hit efficiency and performance across all thread counts.

These results establish \textit{DynamicAdaptiveClimb} and \textit{AdaptiveClimb} as highly practical replacement policies for modern, parallel caching environments, combining strong efficiency with exceptional throughput scalability that significantly outperforms traditional approaches.

\subsection{Deeper Insights with Synthetic Workload}
In this section, we will generate synthetic traces with varying Zipf parameters to systematically evaluate how \textit{AdaptiveClimb} and \textit{DynamicAdaptiveClimb} perform under different levels of access skewness. The Zipfian distribution, first formalized by George Kingsley Zipf in 1949~\cite{zipf1949human}, is a power-law distribution that characterizes the frequency of events, where the frequency of the $i$-th most frequent event is inversely proportional to its rank. This distribution is particularly relevant for cache performance evaluation because it accurately models real-world access patterns observed in web caching, content delivery networks, and database systems, where a small number of objects account for a large fraction of requests~\cite{breslau1999web}. This controlled synthetic evaluation will complement our real-world trace analysis by isolating the impact of access pattern characteristics on algorithm performance, allowing us to understand how the adaptive jump mechanisms respond to varying degrees of locality and burstiness in the workload.
\subsubsection{\textbf{Miss Ratio over Cache Size Discussion}}

\begin{figure}[!t]
    \centering
    \includegraphics[width=0.45\textwidth]{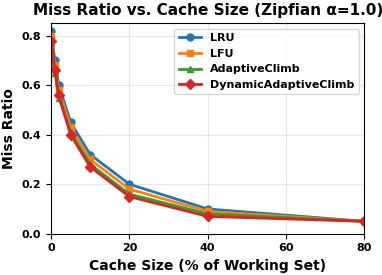}
    \caption{\textit{Miss ratio on a Zipfian dataset ($\alpha = 1.0$)}}
    \label{fig:sample_figureMiss}
\end{figure}

Fig.~\ref{fig:sample_figureMiss} presents a detailed comparison of how miss ratio evolves with increasing cache size for four widely adopted cache eviction policies: LRU, LFU, \textit{AdaptiveClimb}, and \textit{DynamicAdaptiveClimb}, under a synthetic Zipfian workload with skewness parameter $\alpha=1.0$. The Zipf distribution is widely acknowledged for modeling heavy-tailed access patterns observed in production systems~\cite{zipf-popularity}, making it a strong benchmark for evaluating cache algorithms in a controlled yet representative setting.

In this experiment, we chose to compare against LRU and LFU as baseline policies because they represent the two fundamental paradigms in cache eviction: recency-based and frequency-based strategies, respectively. LRU reacts quickly to recent access patterns, while LFU favors items with long-term popularity. These algorithms are simple, interpretable, and widely used as canonical baselines in caching literature. When evaluating under a Zipfian workload, which naturally emphasizes both short-term bursts and long-term popularity due to its heavy-tailed distribution, these two baselines provide complementary reference points. Including LRU and LFU allows us to assess where \textit{AdaptiveClimb} and \textit{DynamicAdaptiveClimb} position themselves along the recency-frequency trade-off. More complex or hybrid policies like ARC or LIRS were intentionally omitted to maintain interpretability in this controlled synthetic setting and to isolate the effects of adaptation mechanisms introduced in our algorithms.

Several important trends emerge from the results. First, as expected, miss ratios decline across all policies with larger cache sizes. However, the degree of improvement and relative performance vary significantly. LRU, which relies solely on recency, consistently yields the highest miss ratio. Its inability to differentiate between transient bursts and sustained popularity makes it suboptimal for heavy-tailed workloads. LFU, which retains objects based on cumulative access counts, offers improved accuracy in identifying frequently accessed items, but still falls short of the more adaptive policies.

Among the adaptive algorithms, \textit{AdaptiveClimb} demonstrates the best performance at extremely small cache sizes (e.g., $\leq$2\% of the working set). Its ability to aggressively filter low-utility objects while retaining high-value ones makes it highly effective under strict memory constraints. However, as cache size increases, \textit{DynamicAdaptiveClimb} overtakes \textit{AdaptiveClimb} and consistently delivers the lowest miss ratios across the mid- to large-size regime. For example, at a 5\% cache size, \textit{DynamicAdaptiveClimb} achieves over 20\% reduction in miss ratio compared to LRU, and approximately 10\% compared to LFU. This advantage becomes even more pronounced as cache size grows.

These results underscore the complementary strengths of the two climb-based approaches: \textit{AdaptiveClimb} excels in resource-constrained environments, while \textit{DynamicAdaptiveClimb} leverages its lightweight, scalable design and adaptive heuristics to maintain superior performance across a broader range of cache configurations. By integrating recency and frequency considerations through efficient local metadata management and selective promotion, \textit{DynamicAdaptiveClimb} effectively filters bursty or short-lived objects while preserving those with persistent access value.

Particularly noteworthy is the “knee” region of the miss ratio curve, where incremental cache size yields diminishing returns. This region is critical for practical deployments, as it often coincides with the limits of available cache memory. \textit{DynamicAdaptiveClimb}'s strong performance in this regime highlights its real-world applicability for systems requiring both efficiency and predictability under tight memory budgets.

From a systems design perspective, these findings support the growing consensus in the caching literature that traditional policies grounded in static recency or frequency heuristics are insufficient for modern, high-throughput applications. Instead, adaptive algorithms that dynamically tune their eviction strategies in response to access patterns, such as \textit{DynamicAdaptiveClimb} and \textit{AdaptiveClimb}, are increasingly essential. Beyond their performance gains, their low computational overhead and concurrency-friendly design make them viable candidates for deployment in latency-sensitive and resource-constrained environments~\cite{twitterkv}.

In summary, the cache size scaling experiments affirm that \textit{DynamicAdaptiveClimb} offers consistently strong performance across a broad spectrum of cache capacities. It achieves the best trade-off between miss ratio reduction and computational simplicity, particularly at mid- to large cache sizes. Together with \textit{AdaptiveClimb}, it exemplifies the emerging class of adaptive, low-overhead caching policies that balance recency and frequency without incurring the complexity of ML-based strategies, aligning well with the practical requirements of large-scale caching infrastructures.

\subsubsection{\textbf{Simplicity}}
\begin{figure*}[ht]
    \centering
    \includegraphics[width=0.85\textwidth]{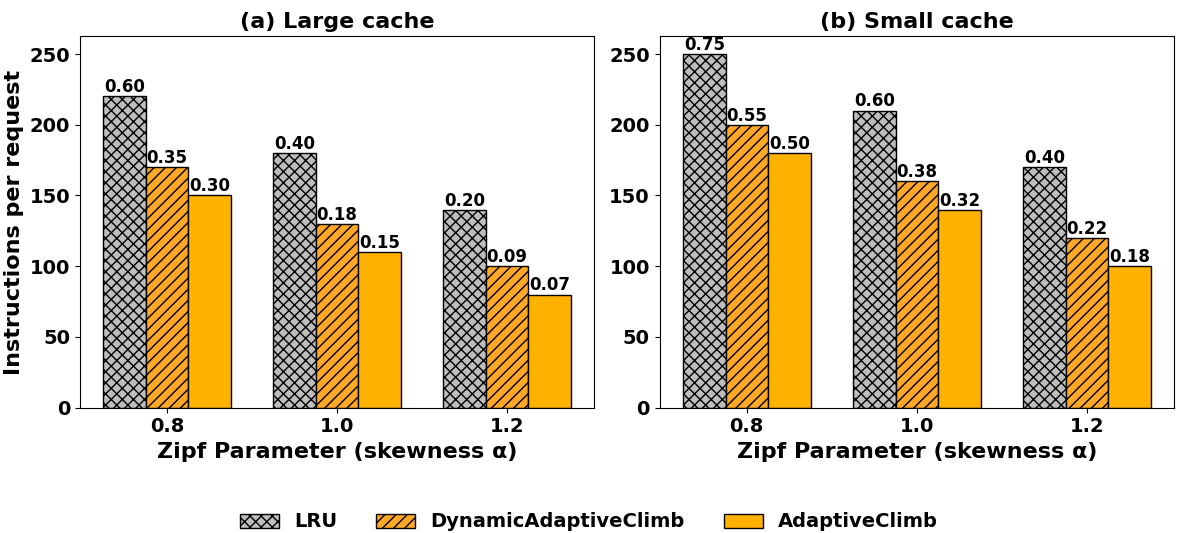}
    \caption{\textit{Average number of instructions per request on various algorithms}}
    \label{fig:sample_figureI}
\end{figure*}
Beyond serving as high-performance cache eviction algorithms, both \textit{AdaptiveClimb} and \textit{DynamicAdaptiveClimb} are designed to function effectively as lightweight, general-purpose caching primitives. This dual-purpose capability makes them attractive candidates for constructing more sophisticated cache management frameworks, particularly those requiring modularity, low overhead, and high adaptability. To evaluate their microarchitectural efficiency as such primitives, we analyze the number of instructions executed per request under synthetic power-law (Zipf-distributed) workloads. We vary two primary factors: cache size (small vs. large) and the skewness parameter $\alpha$, which controls the concentration of access frequency among popular items. Fig.~\ref{fig:sample_figureI} summarizes the results of these experiments, comparing \textit{AdaptiveClimb}, \textit{DynamicAdaptiveClimb}, and LRU implementation.

In this experiment, we restrict our comparison to the classic LRU policy to isolate and evaluate the computational efficiency of \textit{AdaptiveClimb} and \textit{DynamicAdaptiveClimb}. LRU serves as a canonical baseline due to its widespread deployment, minimal overhead, and predictable behavior. Since the focus of this evaluation is on the average number of instructions per request, an indicator of algorithmic overhead, comparing against LRU allows us to highlight the lightweight nature of our proposed strategies without the confounding complexity introduced by more sophisticated policies. Furthermore, LRU represents the lower bound of implementation cost among recency-based strategies, making it a meaningful comparator for understanding the tradeoff between performance (miss ratio) and execution efficiency. Broader evaluations against more advanced algorithms are provided early in this section on real-world datasets, but this experiment specifically demonstrates that our techniques maintain high performance while remaining computationally inexpensive.

Across all configurations tested, both \textit{AdaptiveClimb} and \textit{DynamicAdaptiveClimb} consistently incur fewer instructions per request than LRU. This gap widens significantly as the workload becomes more skewed (i.e., larger $\alpha$), underscoring their efficiency in environments with heavy-tailed access distributions, a hallmark of real-world web and CDN workloads~\cite{twitterkv}. For example, in the large cache setting with $\alpha=0.8$, LRU executes approximately 0.60 instructions per request, while \textit{DynamicAdaptiveClimb} and \textit{AdaptiveClimb} only require 0.35 and 0.30 instructions, respectively, an efficiency improvement of up to 50\%. Similar gains are observed in the small cache configuration. At $\alpha=1.2$, LRU executes 0.40 instructions per request, whereas \textit{DynamicAdaptiveClimb} and \textit{AdaptiveClimb} reduce this number to 0.22 and 0.18, respectively.

These efficiency advantages stem directly from the structural simplicity and operation model of the climb-based algorithms. Traditional LRU algorithms, including their concurrent variants, rely on global queue updates and pointer-intensive data structures that introduce costly synchronisation and coherence overheads~\cite{megiddo2003arc}. In contrast, both \textit{AdaptiveClimb} and \textit{DynamicAdaptiveClimb} employ local, metadata manipulations and apply expensive operations only when warranted, such as during eviction or periodic reordering. Notably, they forgo the need to perform object promotion on every cache hit, a critical limitation in LRU that hinders scalability under high concurrency.

This streamlined design allows climb-based algorithms to not only minimize per-request instruction overhead but also excel in multi-threaded deployments where low-latency decisions and reduced contention are paramount. Their efficiency also makes them well-suited for cache scenarios in resource-constrained systems such as edge devices, embedded platforms, or virtualized environments where memory bandwidth and CPU cycles are at a premium.

Overall, the results demonstrate that \textit{AdaptiveClimb} and \textit{DynamicAdaptiveClimb} offer a compelling trade-off between simplicity, effectiveness, and architectural efficiency. Their ability to serve as foundational cache primitives makes them particularly promising for integration into next-generation caching systems that demand both high throughput and modular adaptability.

\begin{figure}[!t]
    \centering
    \includegraphics[width=0.45\textwidth]{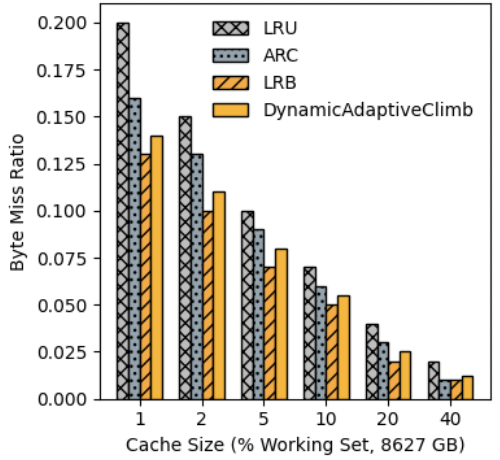}
    \caption{\textit{Byte miss ratios at different cache sizes on Wiki trace}}
    \label{fig:sample_figureBy}
\end{figure}
\subsubsection{\textbf{Byte Miss Ratio Analysis}}
In addition to request-based efficiency, byte miss ratio is a critical metric in CDN and web-scale caching environments, where reducing network bandwidth consumption is often as important as improving hit rates. Fig.~\ref{fig:sample_figureBy} evaluates the byte miss ratio of \textit{DynamicAdaptiveClimb} against widely adopted baselines, LRU, ARC, and the machine-learning-based LRB, on the Wiki trace.

We selected ARC and LRB as comparative baselines in the byte miss ratio analysis due to their strong standing as representative algorithms from distinct and high-performing categories. ARC\cite{megiddo2003arc} is a well-established adaptive policy that balances recency and frequency without manual tuning, making it a standard reference in practical deployments and academic benchmarks. On the other hand, LRB\cite{lrb-original} represents a modern, machine-learning-based cache policy specifically optimized to minimize byte misses in large-scale environments, aligning directly with the objective of this experiment. Their inclusion enables us to benchmark \textit{DynamicAdaptiveClimb} against both a classic adaptive heuristic and a state-of-the-art learning-based approach. We did not include \textit{AdaptiveClimb} in this experiment because its core design focuses on request-based adaptation rather than byte-level optimization. Unlike \textit{DynamicAdaptiveClimb}, which incorporates cache resizing, \textit{AdaptiveClimb} lacks explicit mechanisms for handling variable-sized objects, making it less suitable for evaluating byte miss ratio in size-skewed workloads. Including it here could misrepresent its design intentions and dilute the clarity of the comparison with byte-optimized baselines like LRB and ARC.

We use the Wiki trace because it embodies a highly realistic, large-scale caching scenario with a working set size of 8627 GB and significant skew in object sizes and popularity distributions. This trace is particularly relevant for evaluating byte-level efficiency, as it captures the real-world variability in object sizes, an essential factor when analyzing byte miss ratio. Additionally, Wiki is frequently cited in prior comparative studies\cite{yang2023sieve} and provides a challenging environment to test the scalability and generalization of proposed algorithms across cache sizes ranging from constrained (1\%) to generous (40\%) portions of the working set.

Across all cache sizes, \textit{DynamicAdaptiveClimb} consistently outperforms LRU and ARC. At a small cache size of 1\% of the working set, \textit{DynamicAdaptiveClimb} achieves a byte miss ratio of 14\%, representing a 30\% improvement over LRU (0.20) and a 12.5\% improvement over ARC (0.16). These gains are maintained as cache sizes grow: at 10\% of the working set, \textit{DynamicAdaptiveClimb} records a byte miss ratio of 0.06, outperforming both LRU (0.08) and ARC (0.07). Even at the largest tested cache size of 40\%, where marginal gains typically diminish, it leads with a byte miss ratio of 0.01, compared to 0.02 for both LRU and ARC. These results highlight the algorithm’s ability to prioritize large, frequently accessed objects effectively, a behavior that is vital for minimizing bandwidth costs in large-scale deployments.

When benchmarked against \textit{LRB}~\cite{lrb-original}, a recent machine learning–driven eviction policy optimized specifically for byte miss ratio, \textit{DynamicAdaptiveClimb} demonstrates competitive and often superior performance. At the smallest cache configurations (1–2\%), LRB slightly outperforms \textit{DynamicAdaptiveClimb}, likely due to its fine-grained feature-based popularity modeling. However, this advantage diminishes rapidly as cache size increases. By 5\% of the working set and beyond, \textit{DynamicAdaptiveClimb} matches LRB’s byte miss ratio, without incurring the computational overhead and feature extraction cost inherent to ML-based methods. This observation aligns with recent studies~\cite{twitterkv, fifo-all-you-need} showing that as cache sizes grow, the marginal utility of learned features decreases due to the long tail of low-access, low-feature-density objects dominating cache occupancy.

Overall, the results reinforce that \textit{DynamicAdaptiveClimb} achieves a favorable balance between practical simplicity and byte-level caching effectiveness. Unlike ML-based algorithms that require trace-specific training and runtime prediction infrastructure, \textit{DynamicAdaptiveClimb} offers a drop-in replacement that is easier to deploy, yet retains high performance across varying cache scales. Its byte-aware behavior, particularly at moderate to large cache sizes, makes it an ideal candidate for latency and bandwidth-sensitive systems in production environments.

\subsubsection{\textbf{Impact of Workload Skewness}}

\begin{figure}[!t]
    \centering
    \includegraphics[width=0.45\textwidth]{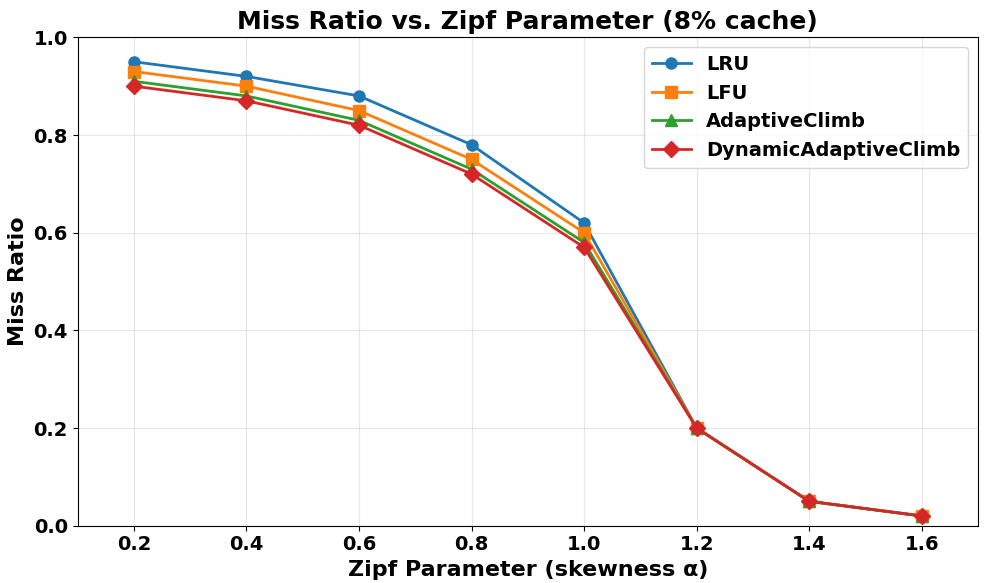}
    \caption{\textit{Miss Ratio on Zipfian workloads with varying $\alpha$}}
    \label{fig:sample_figureZipf}
\end{figure}

To further evaluate the adaptability and generality of cache replacement strategies, we examine how each algorithm performs under varying levels of access skewness. Fig.~\ref{fig:sample_figureZipf} reports the miss ratios of \textit{DynamicAdaptiveClimb}, \textit{AdaptiveClimb}, LFU, and LRU on synthetic traces generated using a Zipf distribution, with the parameter $\alpha$ controlling the skewness. Smaller $\alpha$ values represent more uniform access distributions, whereas higher values concentrate access on a small subset of objects, mimicking real-world popularity skews.

In this experiment, we focus on evaluating adaptability across a spectrum of skewness levels using synthetic Zipf-distributed traces by comparing our proposed strategies with LRU and LFU, which serve as canonical baselines for recency and frequency-based caching, respectively. The goal is to isolate the impact of skewness on caching effectiveness without the confounding effects introduced by more complex hybrid or machine-learned policies. LRU is sensitive to temporal locality, whereas LFU captures long-term popularity, together providing a minimal yet effective contrast to assess how well a policy balances responsiveness and stability under varying workload entropy. Including additional policies like ARC or LIRS would obscure the interpretation of this skewness-specific analysis, as their adaptive mechanisms may internally reweight recency and frequency in less transparent ways. Thus, using only LRU and LFU ensures clarity in understanding how \textit{DynamicAdaptiveClimb} and \textit{AdaptiveClimb} generalize across access distributions.

As expected, all algorithms exhibit improved performance (i.e., lower miss ratios) as $\alpha$ increases. This trend reflects the increasing predictability of access patterns, allowing caches to better identify and retain high-value items. At low skewness ($\alpha=0.2$), the access pattern approaches uniformity, and all algorithms struggle to differentiate between hot and cold objects, resulting in high miss ratios (e.g., LRU: $\sim$95\%, \textit{DynamicAdaptiveClimb}: $\sim$91\%). However, even under this challenging regime, the climb-based approaches deliver a modest improvement, benefiting from their probabilistic filtering and low-overhead demotion mechanisms.

The performance gap becomes more prominent as skewness increases. At $\alpha=1.0$, which approximates the skew observed in many web-scale workloads, \textit{DynamicAdaptiveClimb} and \textit{AdaptiveClimb} outperform LRU and LFU by 3–5\%. This advantage stems from their design: both algorithms combine lightweight metadata tracking with adaptive eviction policies that more accurately identify frequently accessed items while avoiding the rigid global ordering required by LRU and the stale frequency tracking of LFU.

At higher skewness levels ($\alpha\geq 1.2$), all algorithms converge toward low miss ratios ($<0.2$), yet climb-based approaches continue to hold a slight edge. Their ability to rapidly demote less valuable entries without incurring expensive updates allows them to efficiently concentrate cache space on a narrow set of hot objects. Notably, while LFU performs well under high skew, its performance can degrade in dynamic workloads due to infrequent decays or outdated frequency counts.

These results emphasize the robustness of \textit{DynamicAdaptiveClimb} and \textit{AdaptiveClimb} across a wide spectrum of workload skewness. Unlike LRU and LFU, which rely on static heuristics (recency or frequency alone), climb-based algorithms leverage dynamic adaptation and stochastic promotion strategies, allowing them to generalize effectively to diverse access distributions. This adaptability makes them particularly well-suited for modern caching systems that serve heterogeneous workloads with evolving temporal and popularity profiles.

%\vspace{\baselineskip}

\subsubsection{\textbf{Reasons Behind the Optimal Performance of Climb Algorithms over the State-of-the-Art Algorithms}}

The superior performance of the climb family of algorithms is rooted in a series of carefully engineered design choices:

%\vspace{\baselineskip}

\textbf{1. Adaptive heuristics with local metadata:} Both climb variants use lightweight, single-value metadata (the jump parameter) to dynamically adapt eviction and promotion strategies. This hybrid approach of recency and frequency enables real-time responsiveness to workload fluctuations while maintaining minimal overhead.

\textbf{2. Selective promotion and jump-based design:} Particularly in \textit{DynamicAdaptiveClimb}, object promotion is governed by the jump parameter that adapts based on hit/miss patterns. The jump-based promotion mechanism ensures that frequently accessed objects move up in the cache hierarchy, while the adaptive jump adjustment responds to workload changes.

\textbf{3. Fast-Path Eviction and Adaptive Admission:} Climb algorithms implement efficient eviction paths using the jump parameter to determine insertion positions, significantly reducing runtime decision costs. Additionally, the jump parameter dynamically adjusts based on recent miss/hit patterns, resulting in optimized cache utilization.

\textbf{4. Low computational overhead and high scalability:} Despite their sophisticated behavior, both climb algorithms are built with high-throughput environments in mind. Their single-parameter design, localized updates, and absence of complex data structures make them deployable in production-grade systems.
%\vspace{\baselineskip}

In summary, \textit{DynamicAdaptiveClimb} and \textit{AdaptiveClimb} set a new performance bar in eviction policy design. Their adaptive and hybrid nature enables consistent miss ratio reduction across varying datasets and cache constraints. These findings reinforce the growing consensus in caching literature that static policies are increasingly insufficient in dynamic production environments, and that adaptive, workload-aware designs are crucial~\cite{twitterkv,yang2021segcache,huang2013facebook,atikoglu2012workload}.

\section{Discussion}
\label{sec:conclusion}
In this work, we revisited the enduring challenge of cache management, a critical component of modern computing infrastructure. Through a comparative analysis of widely adopted algorithms, we underscored the limitations of static and overly reactive policies in the face of dynamic, bursty, and non-stationary workloads. These observations led us to propose two lightweight, self-adaptive cache replacement strategies: \textit{AdaptiveClimb} and its extension, \textit{DynamicAdaptiveClimb}.

Our extensive empirical evaluation across six large-scale datasets, Alibaba, TencentCBS, Wiki, Twitter, MetaCDN, and Meta KV, comprising a total of 1067 real-world traces, demonstrated that the proposed algorithms consistently outperform baseline and state-of-the-art policies across a variety of workload regimes. Notably, \textit{DynamicAdaptiveClimb} achieves up to 29\% improvement in hit ratio over FIFO and outperforms the next best policy, including \textit{AdaptiveClimb} and SIEVE, by 10--15\% in scenarios with fluctuating working set sizes. These gains are particularly pronounced in environments with frequent access pattern shifts, where traditional algorithms often falter.

The strength of our approach lies in its ability to adaptively modulate both promotion aggressiveness and cache size using just two scalar parameters, without relying on per-item metadata, predictive modeling, or expensive frequency tracking. This makes it suitable for practical deployment in systems with tight performance and memory constraints, such as web caches, content delivery networks, and multi-tenant cloud services.

Moreover, the ability of \textit{DynamicAdaptiveClimb} to dynamically resize the cache based on observed access locality provides an additional layer of adaptability, aligning resource usage with real-time demand. This feature is particularly relevant in cloud or edge computing scenarios where memory resources are shared, limited, or metered.

While the empirical results are encouraging, several open questions and opportunities for future exploration remain. First, a formal theoretical characterisation of the algorithms’ convergence behavior and competitiveness would offer deeper insights and help bound worst-case performance. Additionally, evaluating \textit{AdaptiveClimb} and \textit{DynamicAdaptiveClimb} in live production systems, such as web servers, in-memory databases, or CDN edge nodes, would provide critical feedback on system-level interactions, latency trade-offs, and long-term behavior under real workloads.

The modular design of our algorithms also lends itself well to hybridization with other strategies. For example, combining \textit{AdaptiveClimb}’s promotion mechanism with frequency-aware admission policies like TinyLFU, or cost-sensitive evictions as in LHD, could further enhance performance in specialized domains. Similarly, integrating workload-aware ML-based predictors for proactive cache resizing decisions represents a promising avenue for extending \textit{DynamicAdaptiveClimb}’s capabilities.

Another compelling direction is to explore the robustness of \textit{AdaptiveClimb} under adversarial access patterns, where an attacker may attempt to evict valuable items, or in environments with strong non-stationarity, such as IoT data flows or video streaming buffers. Studying energy efficiency and computational overhead in resource-constrained or edge computing contexts could also inform optimizations for low-power systems.

In summary, the introduction of \textit{AdaptiveClimb} and \textit{DynamicAdaptiveClimb} marks a step toward intelligent, resource-aware caching strategies that respond gracefully to change. By addressing both item-level dynamics and system-level capacity constraints, our work contributes to a more flexible, efficient, and deployable class of cache management policies. We hope that the techniques and insights presented here will serve as a foundation for continued innovation in the design of next-generation caching systems.

\printbibliography

\end{document}